\documentclass[runningheads, anonymous]{lncs}
\PassOptionsToPackage{draft}{hyperref}
\usepackage{float, enumerate, xcolor, algorithm, algorithmic, tabto, balance, amsmath, graphicx, epstopdf, hyperref, multirow, dblfloatfix, textcomp, listings, eurosym, lstautogobble, multirow, tikz} %textcomp for copyright symbol
\usepackage{amsfonts}
\usepackage{caption}

\usepackage{amssymb}
\usepackage{booktabs}
\usepackage{url}
\usepackage{amssymb}
\usepackage[numbers,sort&compress]{natbib}
\usepackage[flushleft]{threeparttable}
\usepackage{subfigure}
\usetikzlibrary{arrows,backgrounds,positioning}
\usepackage{chngcntr}
\usepackage[overload]{empheq}
\usepackage{soul}
\usepackage[multiple]{footmisc} % for comma between two footnotes at the same place
\usepackage[most]{tcolorbox} %for drawing border around an image
\usepackage[title]{appendix}

\newcommand{\ie}{{i.e.,~}}
\newcommand{\eg}{{e.g.,~}}

\newcommand{\etc}{{,~etc. }}
\newcommand{\etal}{{et~al. }}

\newcommand{\bigcdot}{\boldsymbol{\cdot}}

\newcommand{\ankit}[1]{\textcolor{black}{#1}}

\newcommand{\cans}[1]{\textcolor{black}{#1}}
\newcommand{\canss}[1]{\textcolor{black}{#1}}

\usepackage{array}
\newcolumntype{L}{>{\centering\arraybackslash}m{.5\columnwidth}|}
\newcommand{\fixcolumn}[1]{\multicolumn{1}{m{.5\columnwidth}|}{#1}}

\usepackage{url}

\usepackage{pifont}% http://ctan.org/pkg/pifont
\newcommand{\cmark}{\ding{51}} %BOLD CHECK
\newcommand{\xmark}{\ding{55}} %BOLD CROSS
\usetikzlibrary{arrows,automata}

\graphicspath{ {./images/} }

\begin{document}
%\title{Detecting Covert Cryptomining using Hardware-assisted Profiling}

% This is samplepaper.tex, a sample chapter demonstrating the
% LLNCS macro package for Springer Computer Science proceedings;
% Version 2.20 of 2017/10/04
%
% Used for displaying a sample figure. If possible, figure files should
% If you use the hyperref package, please uncomment the following line
% to display URLs in blue roman font according to Springer's eBook style:
% \renewcommand\UrlFont{\color{blue}\rmfamily}

	%
	\title{Detecting Covert Cryptomining using HPC}
	
	\author{\canss{
		Ankit~Gangwal \inst{1} \and %\thanks{Corresponding author} \and
		Samuele~Giuliano~Piazzetta \inst{2} \and
		Gianluca~Lain \inst{2} \and
		Mauro~Conti \inst{3}}
	}
	\institute{\canss{
		TU Delft, Netherlands
		\and
		ETH Z\"{u}rich, Switzerland
		\and
		University of Padua, Italy\\
		\email{\{a.gangwal@tudelft.nl, spiazzetta@student.ethz.ch, gilain@student.ethz.ch, conti@math.unipd.it\}}}
	}
	\authorrunning{\canss{A. Gangwal et al.}}
	
%				\author{Anonymous}
%				\institute{Anonymous}
%				\authorrunning{Anonymous}
	\maketitle              % typeset the header of the contribution
	\begin{abstract}
			Cybercriminals have been exploiting cryptocurrencies to commit various unique financial frauds. Covert cryptomining - which is defined as an unauthorized harnessing of victims' computational resources to mine cryptocurrencies - is one of the prevalent ways nowadays used by cybercriminals to earn financial benefits. Such exploitation of resources causes financial losses to the victims.%Covert cryptomining is one of the prevalent ways nowadays used by cybercriminals to earn financial benefits by harnessing the computational resources of the victims. 
		\par
		In this paper, we present our efficient approach to detect covert cryptomining \cans{on users' machine}. Our solution is a generic solution that, unlike currently available solutions to detect covert cryptomining, is not tailored to a specific cryptocurrency or a particular form of cryptomining. In particular, we focus on the core mining algorithms and utilize Hardware Performance Counters (HPC) to create clean signatures that grasp the execution pattern of these algorithms on a processor. We built a complete implementation of our solution employing advanced machine learning techniques. We evaluated our methodology on two different processors through an exhaustive set of experiments. In our experiments, we considered all the cryptocurrencies mined by the top-10 mining pools, which collectively represent the largest share of the cryptomining market. Our results show that our classifier can achieve a near-perfect classification with samples of length as low as five seconds. Due to its robust and practical design, our solution can even adapt to zero-day cryptocurrencies. Finally, we believe our solution is scalable and can be deployed to tackle the uprising problem of covert cryptomining.
		
	\end{abstract}

\section{Introduction}
Cryptomining, or simply mining, is a process of validating and adding new transaction in the blockchain digital ledger for various cryptocurrency. It is an essential process to keep most of the cryptocurrencies running. Typically, mining is a resource-intensive process that continuously performs heavy computations. Upon successful mining, miners receive newly generated cryptocoins as their remuneration. Usually, newer cryptocurrencies tend to pay a higher reward. Some cryptocurrencies, such as Monero, make mining feasible on the web-browsers that enable even layman users to participate in mining.
\par
After the success of Bitcoin~\cite{nakamoto2008bitcoin}, many alternative cryptocurrencies (altcoins) have been introduced to the market. \ankit{At the time of writing, there are over 2000 active cryptocurrencies~\cite{coinmarketcap}.} The massive number of cryptocurrencies raises an enormous demand for mining. This demand continues to remain huge because mining, as mentioned before, is an inevitable operation to keep these virtual currency systems running. Such an immense demand for mining has attracted cybercriminals~\cite{h1, h2} to earn financial gains, who have already been exploiting cryptocurrencies to perform several types of financial crimes, \eg ransomware~\cite{conti2018economic}. 
\par
\textit{Motivation:} A genuine miner has to make an investment in hardware and bear the significant cost of electricity to run the mining hardware as well as cooling facilities~\cite{a1}. Nevertheless, mining is not beneficial on personal expenditure (mainly, on electricity) unless mining is performed with specialized hardware~\cite{a2}. However, mining can be very profitable if it is performed with ``stolen'' resources, \eg through covert cryptomining, or simply cryptojacking. Cryptojacking is defined as an unauthorized use of the computing resources on a computer, tablet, mobile phone, or connected home device to mine cryptocurrencies. 
\par
\ankit{Cybercriminals have made several ingenious attempts to spread cryptojackers in the form of malware~\cite{webcobra}, malicious browser extensions~\cite{facexworm}\etc by exploiting vulnerability~\cite{rtorrent}, compromising third-party plug-ins~\cite{library}, maneuvering misconfigurations~\cite{misconfiguration}, taking advantage of web-based hosting service~\cite{github}, and so on. To evade intrinsic detection techniques (\eg processor's usage), some cryptojackers suspend their execution when the victim is using the computer~\cite{minergate}, use ``pop-under'' windows to keep mining for a comparatively longer duration~\cite{pop_under}, and utilize legitimate processes of the operating system to mine~\cite{badshell}. Moreover, merely monitoring CPU load\etc is an ineffective strategy because of both false positives and false negatives~\cite{konoth2018minesweeper}.}
\par
To further aggravate the situation, cryptocurrency mining service (\eg Coinhive~\cite{coinhive}, Crypto-Loot~\cite{cryptoloot}) easily integrate into websites to monetize the computational power of their visitors. In fact, cryptojacking attacks exceeded ransomware attacks in 2018 and affected five~times more systems as compared to ransomware~\cite{cryptojacking_compare}. \cans{According to Symantec's report~\cite{cans_symantec_report}, almost double cryptominers were detected on consumer machines as compared to enterprise machines between October 2017 and February 2018 while the same volume of cryptominers was detected on consumer and enterprise machines between March 2018 and July 2018. Kaspersky's report~\cite{cans_kaspersky_report} shows that the total number of internet users who encountered cryptominers rose from 1.9 million in 2016-2017 to 2.7 million in 2017-2018. IBM X-Force Threat Intelligence Index 2019~\cite{cans_ibm_report} estimates that cryptojacking attacks increased by more than 4-times ($\sim$450\%) from Q1 2018 to Q4 2018. SonicWall researchers~\cite{cans_sonicwall_report} reported that cryptojacking attackers made 52.7 million cryptojacking hits during the first half of 2019.} Such exploitation of the computational resources causes financial damage - primarily in the form of increased\footnote{A machine consistently performs heavy computations while it does cryptomining, which, in turn, continuously draws electricity.} electricity bills - to the victims, who often discover the misuse when the damage has already been done. %Additionally, prolonged mining on an incompatible device may also damage the hardware~\cite{and1}. 
\par
On another side, the current state of cryptomining has been consuming a vast amount of energy. As a representative example, Bitcoin~Energy~Consumption~Index was created to provide insight into this amount with respect to Bitcoin, Bitcoin network consumes electricity close to the total demand by Iraq, and a single Bitcoin transaction requires nearly 2.7~times the electrical energy consumed by 100,000~transactions on the VISA~network~\cite{beci}. Moreover, a recent study~\cite{mora2018bitcoin} has suggested that ``Bitcoin usage could alone produce enough $CO_2$ emissions to push warming above $2\,^{\circ}\mathrm{C}$ within less than three decades.'' The current situation would further worsen with illegal/unauthorized/covert cryptomining. Finally, the abundance of the active cryptocurrencies raises the demand for a generic solution to detect covert cryptomining that does not focus on a particular cryptocurrency.
\par
\textit{Contribution:} In this paper, we focus on detecting covert cryptomining \cans{on users' machine}. The major contributions of this paper are as~follows:
\begin{enumerate}
	\item We propose an efficient approach to detect covert cryptomining. In particular, our approach uses HPC to profile the core of the mining process, \ie the mining algorithms, on a given processor to accurately identify cryptomining in real-time. We designed our solution to be a generic one, \ie it is not tailored to a particular cryptocurrency or a specific form of cryptomining.
		
	\item We exhaustively assess the quality of our proposed approach. To this end, we designed six different experiments: (1)~\textit{binary}~classification; (2)~\textit{currency} classification; (3)~\textit{nested} classification; (4)~\textit{sample length}; (5)~\textit{feature relevance}; and (6)~\textit{unseen miner programs}. For a thorough evaluation, we considered eleven distinct cryptocurrencies in our experiments. Our results show that our classifier can accurately classify cryptomining activities.  
	
	\item In the spirit of reproducible research, we make our collected datasets and the code publicly available\footnote{\canss{spritz.math.unipd.it/projects/cryptojackers/}}. %spritz.math.unipd.it/projects/cryptojackers/
	%Link hidden to preserve anonymity
\end{enumerate}

\par
\textit{Organization:} The remainder of this paper is organized as follows. Section~\ref{section:related_work} presents a summary of the related works. We explain our system's architecture in Section~\ref{section:system_architecture} and discuss its evaluation in Section~\ref{section:evaluation}. Section~\ref{section:discussion} addresses the potential limitations of our solution. Finally, Section~\ref{section:conclusion} concludes the paper.

\section{Related works}
\label{section:related_work}
HPC are special-purpose registers in modern microprocessors that count and store hardware-related activities. These activities are commonly referred to as hardware \textit{events\footnote{\cans{An \textit{event} is defined as a countable activity, action, or occurrence on a device.}}}. HPC are often used to conduct low-level performance analysis and tuning. HPC-based monitoring has very low-performance overhead, which makes it suitable even for latency-sensitive systems. Several works have shown the effectiveness of using HPC to detect generic malware~\cite{demme2013feasibility, yuan2011security, Wang:HPC}, kernel-level rootkits~\cite{wang2013numchecker}, side-channel attacks~\cite{chiappetta2016real}, unauthorized firmware modifications~\cite{wang2015confirm}\etc 
\par
\ankit{A general-purpose process classification may distinguish a browser application from a media player or one browser application from another browser application. In the former case, the nature of the applications is different while both the applications in the latter case have the same nature and perform the same operation of rendering pages. Cryptominers have the same nature (of mining), but they essentially perform very different underlying operations due to different proof-of-works, and they also require different compute resources (\eg BTC\footnote{\canss{To refer to different cryptocurrencies, we use their standard ticker symbol. See \tablename{~\ref{table:mining_algo_miners}} for acronyms and their corresponding cryptocurrencies.}} mining is processor-oriented while XMR mining is memory-oriented). Hence, a comparison of our work with the general-purpose process classification methods falls out of the scope of this paper.}
\par
\ankit{On another side, there are limited number of works on detecting cryptomining. Bonneau~\etal \cite{bonneau2015sok} discuss open research challenges of various cryptocurrencies and their mining. Huang~\etal \cite{huang2014botcoin} present a systematic study of Bitcoin mining malware and have shown that modern botnets tend to do illegal cryptomining. %Eskandari~\etal \cite{eskandari2018first} present a survey of in-browser cryptomining. 
Gangwal~\etal \cite{gangwal2019cryptomining} use magnetic side-channel to detect cryptomining. Other works~\cite{konoth2018minesweeper, liu2018novel, wang18esorics, rauchberger2018other, ruth2018digging} focus particularly on browser-based mining. However, only a limited number of cryptocurrencies can be mined in the web-browsers. MineGuard~\cite{tahir2017mining} focuses on detecting cryptomining operations in the cloud~infrastructure.}
\par
\ankit{Our work is different from the state-of-the-art on the following dimensions: (1)~our proposed solution is a generic solution that is not tailored to a particular cryptocurrency or a specific form (e.g., browser-based) of cryptomining on computers; and (2)~we tested our solution against all the cryptocurrencies mined~by the top-10 mining pools, which collectively represent the largest portion of the cryptomining business.}

\section{System architecture}
\label{section:system_architecture}
\cans{We elucidate the key concept behind our approach in Section~\ref{section:core_concept}, our data collection phase in Section~\ref{section:data_collection}, selection of cryptocurrencies in Section~\ref{section:cryptocurrencies_miners}, and our classifier's design in Section~\ref{section:classifier_design}.}

\subsection{Fundamental intuition of our approach}
\label{section:core_concept}
\begin{figure*}[!htbp]
	\vspace{-1.5em}
	\centering
	\subfigure{
		\centering
		\includegraphics[trim = 10mm 145mm 10mm 2mm, clip, width=.7\linewidth]{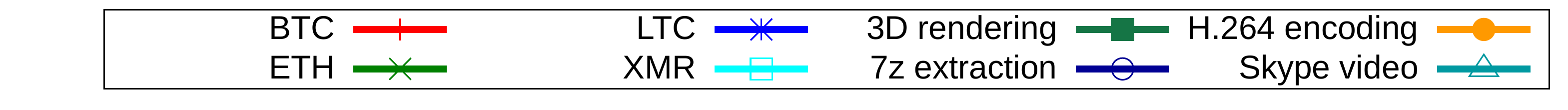}
	}
	\addtocounter{subfigure}{-1}
	\vspace{-0.5em}
	
	\subfigure[\cans{L1-dcache-loads}]{
		\centering
		\includegraphics[trim = 2mm 2mm 2mm 2mm, clip, width=.41\linewidth]{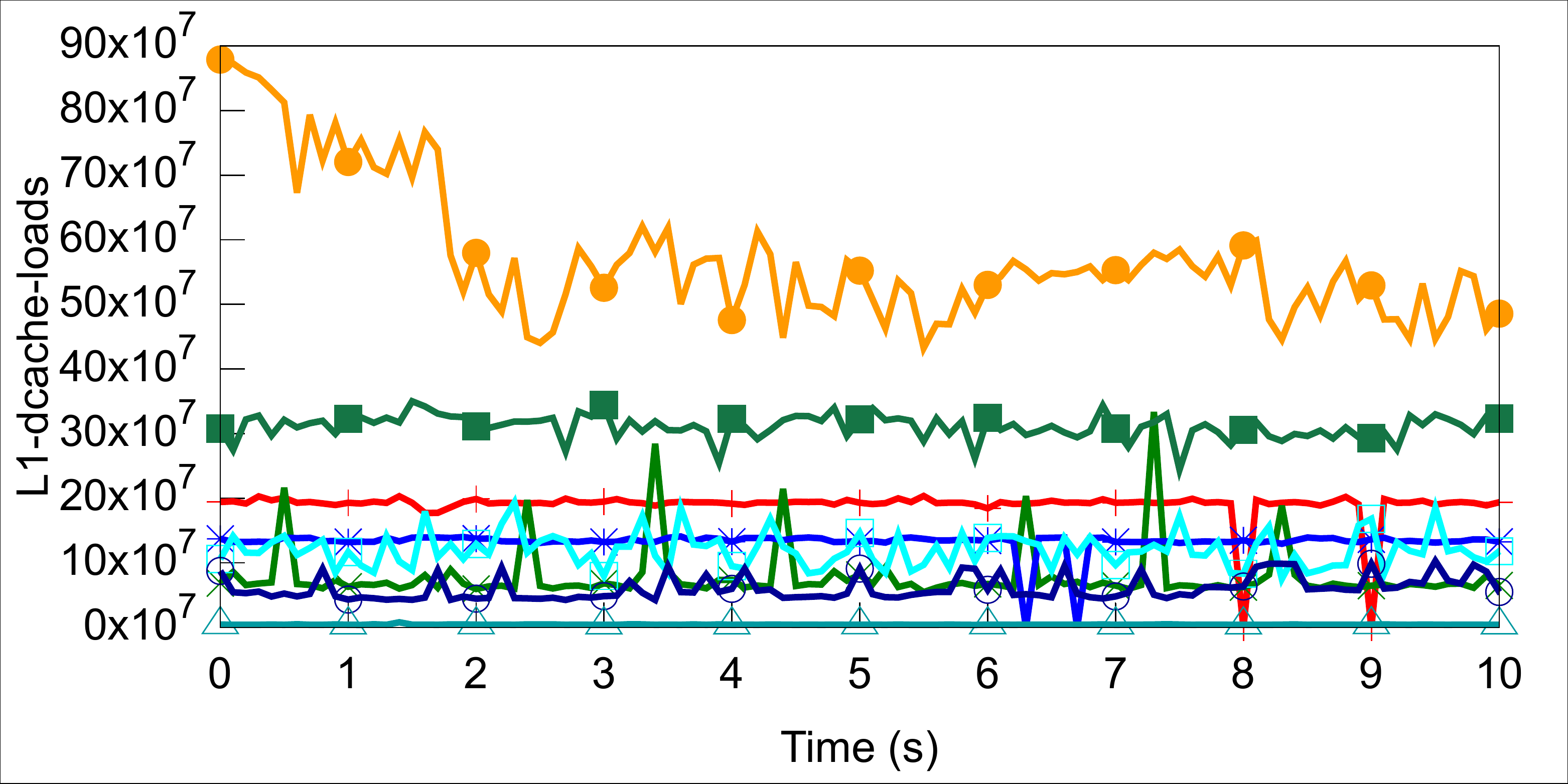}
	}
	\subfigure[\cans{L1-dcache-loads-misses}]{
		\centering
		\includegraphics[trim = 2mm 2mm 2mm 2mm, clip, width=.41\linewidth]{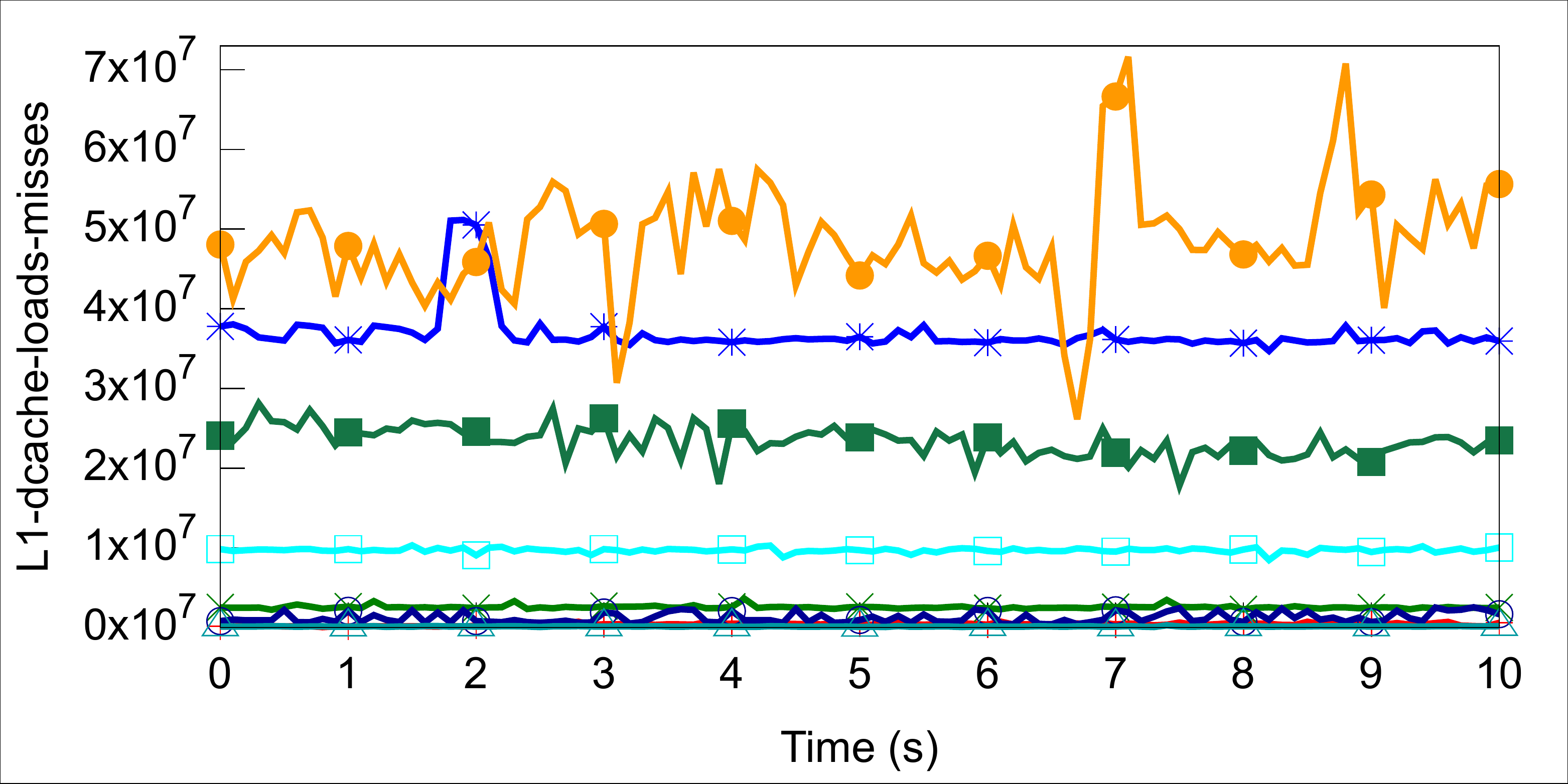}
	}	
	\vspace{-0.5em}

	\subfigure[\cans{Instructions}]{
		\centering
		\includegraphics[trim = 2mm 2mm 2mm 2mm, clip, width=.41\linewidth]{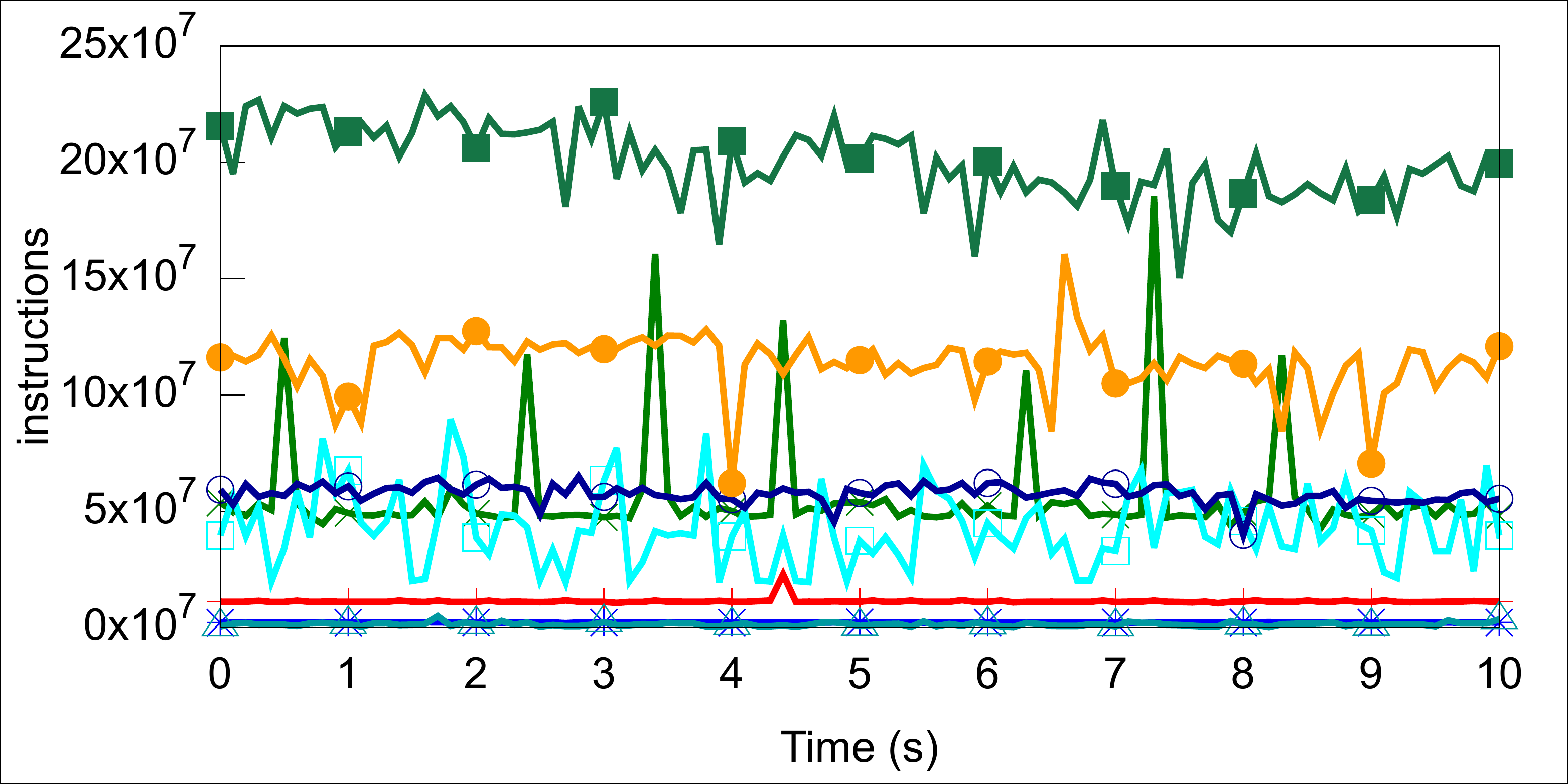}
	}
	\subfigure[\cans{dTLB-loads}]{
		\centering
		\includegraphics[trim = 2mm 2mm 2mm 2mm, clip, width=.41\linewidth]{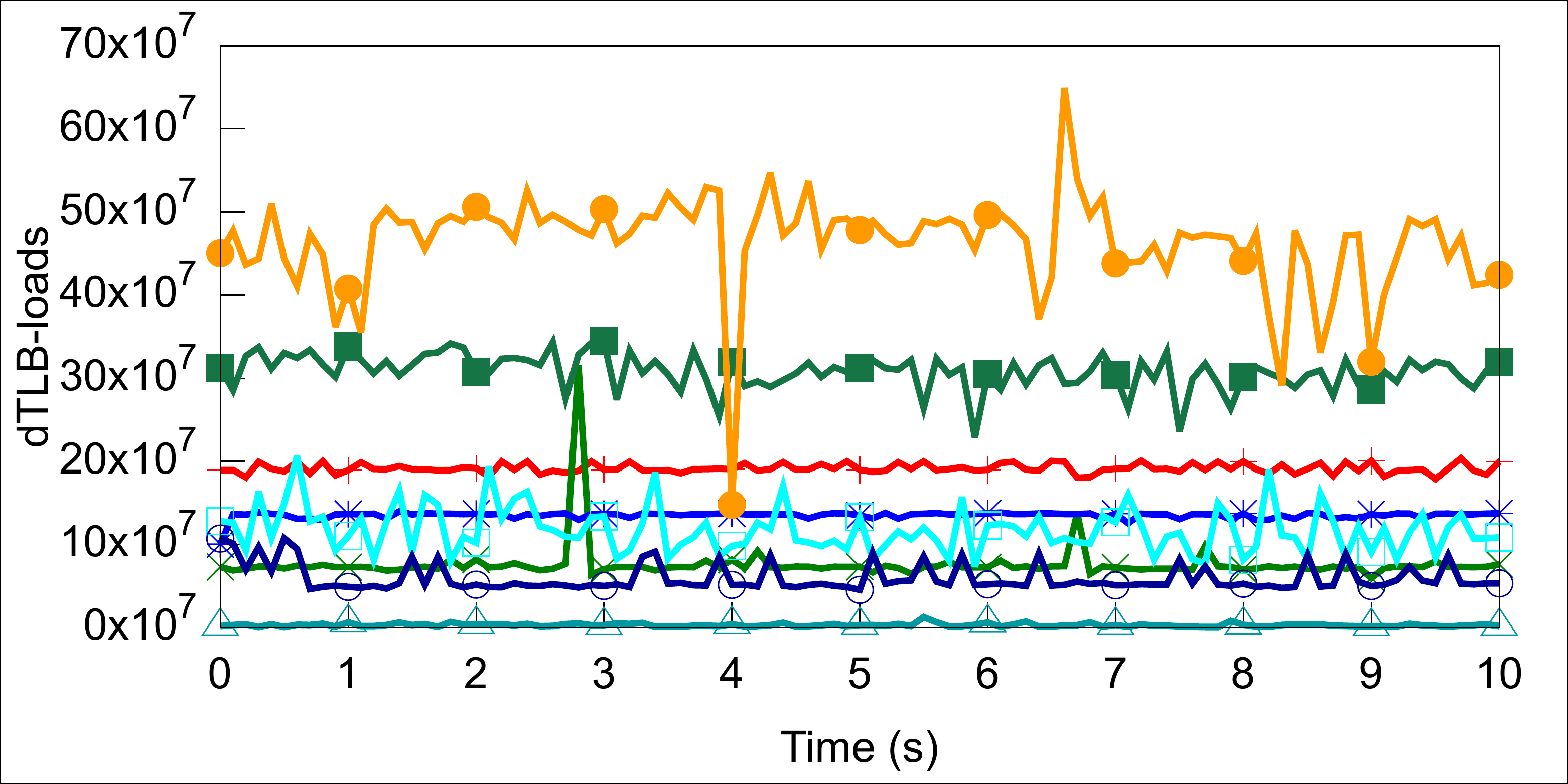}
	}	
	\vspace{-0.5em}
	\subfigure[\cans{Cache-misses}]{
		\centering
		\includegraphics[trim = 2mm 2mm 2mm 2mm, clip, width=.41\linewidth]{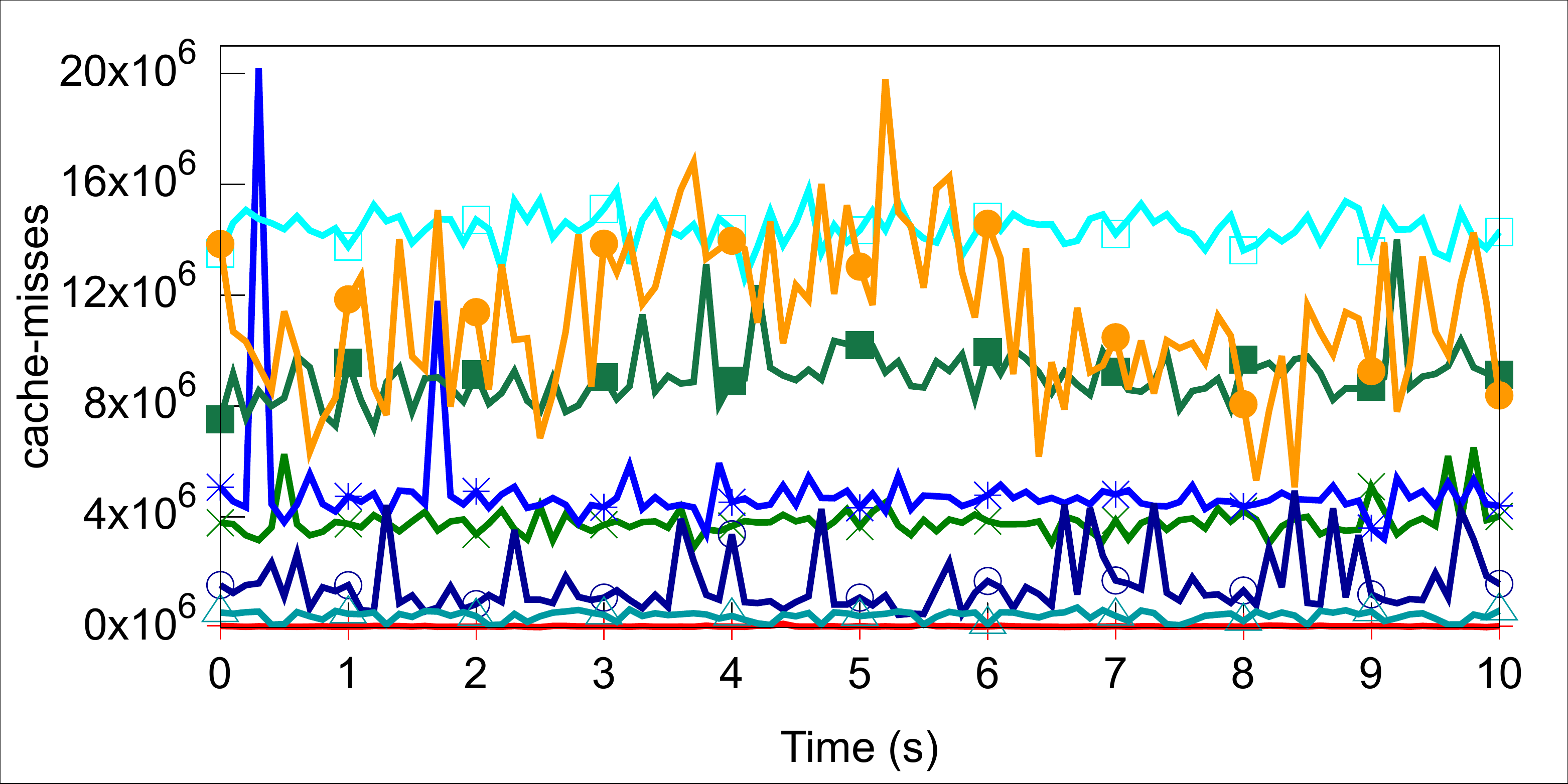}
	}		
	\subfigure[\cans{Context-switches}]{
		\centering
		\includegraphics[trim = 2mm 2mm 2mm 2mm, clip, width=.41\linewidth]{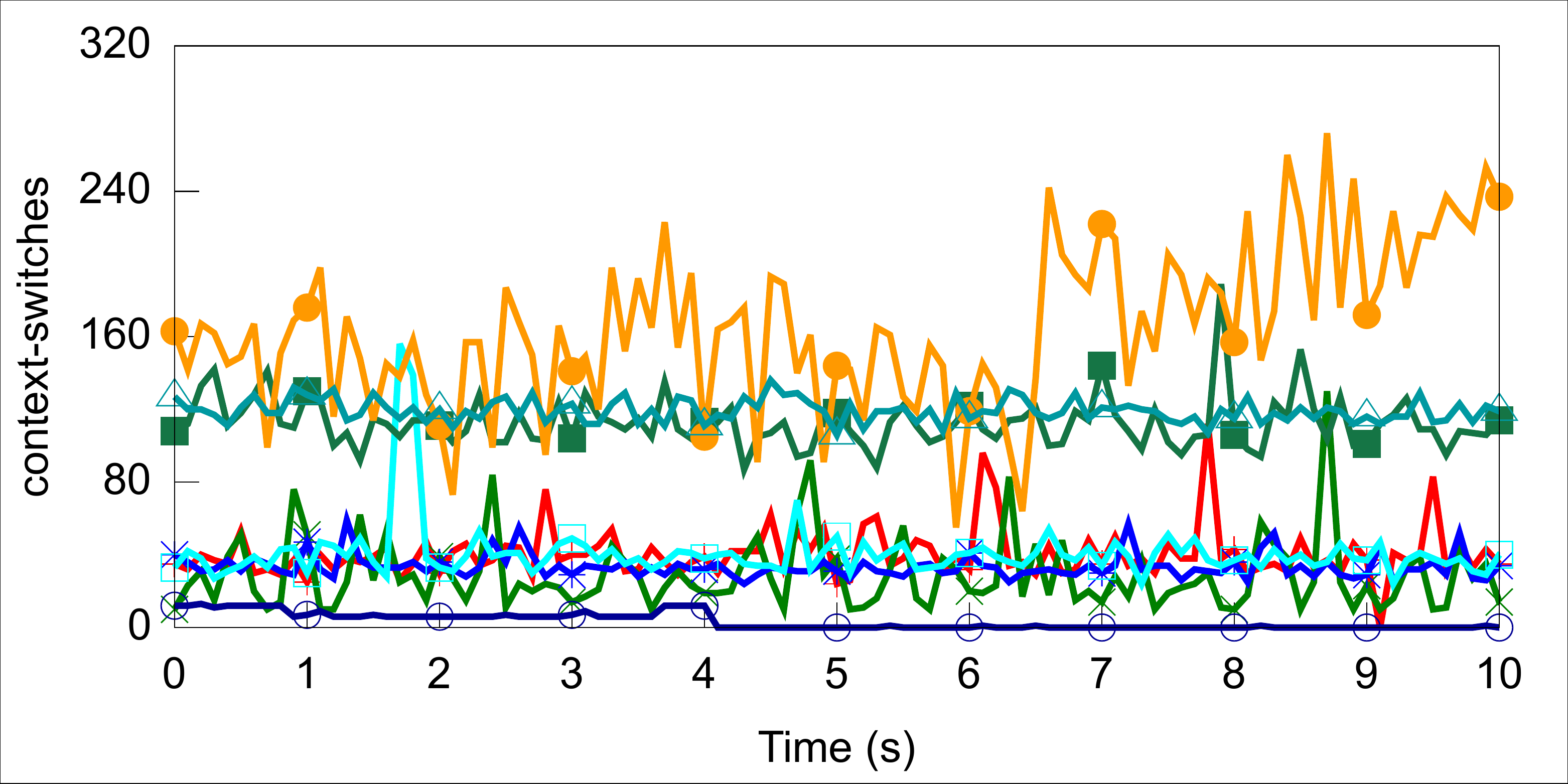}
	}	
	\caption{A representative example of variation in \textit{events} while mining different cryptocurrencies and performing some common user-tasks. HPC were polled every 100ms. The line-points in the graphs do not represent data points and are merely used to make lines distinguishable.}
	\label{figure_pattern}
	\vspace{-2em}
\end{figure*}

The task of cryptomining requires a miner to run the core Proof-of-Work~(PoW\footnote{We use the term ``PoW'' to represent different consensus algorithms.}) algorithm repetitively to solve the cryptographic puzzle. At a coarse-grained level, some PoW algorithms are processor-oriented (\eg BTC) while some are memory-oriented (\eg XMR) due to their underlying design. At a fine-grained level, each PoW algorithm has its own unique mathematical/logical computations (or, in other words, the sequence of operations). Thus, each algorithm upon execution affects some specific \textit{events} more as compared to other \textit{events} on the processor. Consequently, when an algorithm is executed several times repetitively, the ``more'' affected\ \textit{events} outnumber the other~-~relatively under affected~-~\textit{events}. It means that a discernible signature can be built using the relevant \textit{events} for a PoW algorithm. As a representative example, \figurename{~\ref{figure_pattern}} depicts the variation in \textit{events} while mining different cryptocurrencies and performing some common user-tasks. LTC, for instance, shows a more erratic pattern in cache-misses as compared to the other \textit{events} that are affected during LTC mining. On the other hand, a Skype video call has more disparity in context-switches.
\par 
%Finally, in the abundance of active cryptocurrencies, any solution to detect covert cryptomining that focuses on a particular cryptocurrency may not be suitable to tackle the present situation, which raises the demand for a generic solution.
%Several new cryptocurrencies are introduced to the market every day while only a few survive in the business competition. Thus, any solution to detect covert cryptomining that focuses on a particular cryptocurrency or a specific form of cryptomining may not be suitable to tackle such a fast-changing situation.
In practice, there is a finite number of PoW algorithms upon which cryptocurrencies are established. So, we concentrate on the mining algorithms instead of individual currency in our solution. %The primary advantage of our approach is that the signature built for an algorithm would be able to identify even polymorphic, metamorphic, and heavily obfuscated implementations of that algorithm because the core PoW algorithm - that we profile in our solution - remains the same. 
To this end, we use supervised machine learning (cf. Section~\ref{section:classifier_design}) to construct signatures and build~our classifier.
\par
On another side, an adversary may attempt to circumvent such signature-based detection in the following ways: (1) by controlling/limiting the mining; or (2) by neutralizing the signatures. Limiting the mining would reduce the hashing rate, which would indeed make the mining less profitable. Whereas, to neutralize the signatures, the adversary has to succeed in two main hurdles. First, the adversary must have to find those computation(s) that only changes those \textit{events} that are unrelated to the PoW algorithm. Second, the adversary must have to run these computation(s) in parallel to the PoW algorithm, which would again hamper the hashing rate, and thus the profit. In this work, we make a practical assumption that the attacker wants to maximize the profit and does not want to lose the computation cycles (hashing rate).

%Since the miners have to adhere to and repeatedly execute the core PoW algorithm used by a cryptocurrency, the pattern of the magnetic field emitted while mining a cryptocurrency is distinct as well as consistent. \ankit{For the same reason, even a smaller signature database is sufficient for reasonable classification results.}

\subsection{Data collection}
\label{section:data_collection}	
To better explain our work, we first describe what data we collect and how we collect it. We used the \textit{perf}~\cite{perf} tool to profile the processor's \textit{events} using HPC. In particular, we focus on hardware\footnote{Basic events, measured by Performance Monitoring Units (PMU).} \textit{events} (\eg branch-misses), software\footnote{Measurable by kernel counters.} \textit{events} (\eg page-faults), and hardware cache\footnote{Data- and instruction-cache hardware events.} \textit{events} (\eg cache-misses) on CPU as the mining processes - depending on their design - require different type of resources. %are resource intensive. 
We profiled each program of both positive (mining) and negative (non-mining) class individually and collected a total 50 samples per program. Each sample consists of recordings of 28~\textit{events} (described in \tablename{~\ref{table:HPC_events}}) for 30~seconds with a sampling rate of 10Hz, which means that each sample comprises 300~readings of 28~\textit{events}, \ie 8400~readings. To obtain clean signatures: (1) we profiled each program in its stable stage, \ie omitting the bootstrapping phase; and (2) restarted the system to remove any trace of the previous sample.
\begin{table}[!htbp]
	\vspace{-2em}
	\centering
	\caption{\cans{The \textit{events} that we monitor using HPC. Here, HW = hardware,
		SW = software, and HC = hardware cache \textit{event}.}} 
	\label{table:HPC_events}
	\resizebox{.98\columnwidth}{!}
	{

\begin{tabular}{|l|l|l|l|l|l|}
	\hline
	\textbf{\textbf{Event}} & \textbf{\textbf{Type}} & \textbf{\textbf{Description}}                                                                         & \textbf{\textbf{Event}} & \textbf{\textbf{Type}} & \textbf{\textbf{Description}}                                                                        \\ \hline
	branch-instructions     & HW                     & N. of retired branch instructions.                                                                    & iTLB-load-misses        & HC                     & \begin{tabular}[c]{@{}l@{}}N. of instruction fetches that\\ missed instruction TLB.\end{tabular}     \\ \hline
	branch-load-misses      & HW                     & N. of Branch load misses.                                                                             & iTLB-loads              & HC                     & \begin{tabular}[c]{@{}l@{}}N. of instruction fetches that\\ queried instruction TLB.\end{tabular}    \\ \hline
	branch-loads            & HW                     & N. of Branch load accesses.                                                                           & L1-dcache-load-misses   & HC                     & \begin{tabular}[c]{@{}l@{}}N. of load misses at L1 data\\ cache.\end{tabular}                        \\ \hline
	branch-misses           & HW                     & \begin{tabular}[c]{@{}l@{}}N. of mispredicted branch\\ instructions.\end{tabular}                     & L1-dcache-loads         & HC                     & N. of loads at L1 data cache.                                                                        \\ \hline
	bus-cycles              & HW                     & \begin{tabular}[c]{@{}l@{}}N. of bus cycles, which can be\\ different from total cycles.\end{tabular} & L1-dcache-stores        & HC                     & N. of stores at L1 data cache.                                                                       \\ \hline
	cache-misses            & HC                     & N. of cache misses.                                                                                   & LLC-load-misses         & HC                     & \begin{tabular}[c]{@{}l@{}}N. of load misses at the last\\ level cache.\end{tabular}                 \\ \hline
	cache-references        & HC                     & N. of cache accesses.                                                                                 & LLC-loads               & HC                     & \begin{tabular}[c]{@{}l@{}}N. of loads at the last level\\ cache.\end{tabular}                       \\ \hline
	context-switches        & SW                     & N. of context switches.                                                                               & LLC-store-misses        & HC                     & \begin{tabular}[c]{@{}l@{}}N. of store misses at the last\\ level cache.\end{tabular}                \\ \hline
	cpu-migrations          & SW                     & \begin{tabular}[c]{@{}l@{}}N. of times the process has\\ migrated.\end{tabular}                       & LLC-stores              & HC                     & \begin{tabular}[c]{@{}l@{}}N. of stores at the last level\\ cache.\end{tabular}                      \\ \hline
	dTLB-load-misses        & HC                     & N. of load misses at data TLB.                                                                        & mem-loads               & HC                     & N. of memory loads.                                                                                  \\ \hline
	dTLB-loads              & HC                     & N. of load hits at data TLB.                                                                          & mem-stores              & HC                     & N. of memory stores.                                                                                 \\ \hline
	dTLB-store-misses       & HC                     & N. of store misses at data TLB.                                                                       & page-faults             & SW                     & N. of page faults.                                                                                   \\ \hline
	dTLB-stores             & HC                     & N. of store hits at data TLB.                                                                         & ref-cycles              & HW                     & \begin{tabular}[c]{@{}l@{}}N. of total cycles; not affected by\\ CPU frequency scaling.\end{tabular} \\ \hline
	instructions            & HW                     & N. of retired instructions.                                                                           & task-clock              & SW                     & \begin{tabular}[c]{@{}l@{}}The clock count specific to the\\ task that is running.\end{tabular}    \\ \hline
\end{tabular}
	}
\vspace{-1em}
\end{table}
%\textit{branch-instructions, branch-load-misses, branch-loads, branch-misses, bus-cycles, cache-misses, cache-references, context-switches, cpu-migrations, dTLB-load-misses, dTLB-loads, dTLB-store-misses, dTLB-stores, instructions, iTLB-load-misses, iTLB-loads, L1-dcache-load-misses, L1-dcache-loads, L1-dcache-stores, LLC-load-misses, LLC-loads, LLC-store-misses, LLC-stores, mem-loads, mem-stores, page-faults, ref-cycles, task-clock}

%\begin{table}[H]
%	\centering
%	\caption{The events that we monitor using HPC}
%	\label{table:HPC_events}
%	\resizebox{.98\columnwidth}{!}
%	{
%		\begin{tabular}{|l|l|l|l|}
%			\hline
%			branch-instructions & context-switches  & iTLB-load-misses      & LLC-store-misses \\ \hline
%			branch-load-misses  & cpu-migrations    & iTLB-loads            & LLC-stores       \\ \hline
%			branch-loads        & dTLB-load-misses  & L1-dcache-load-misses & mem-loads        \\ \hline
%			branch-misses       & dTLB-loads        & L1-dcache-loads       & mem-stores       \\ \hline
%			bus-cycles          & dTLB-store-misses & L1-dcache-stores      & page-faults      \\ \hline
%			cache-misses        & dTLB-stores       & LLC-load-misses       & ref-cycles       \\ \hline
%			cache-references    & instructions      & LLC-loads             & task-clock       \\ \hline
%		\end{tabular}
%	}
%\end{table}

\par
For the positive class, we profiled a total of 11 cryptocurrencies discussed in Section~\ref{section:cryptocurrencies_miners}. As the representatives of negative class, we chose: 3D rendering; \textit{7z} archive extraction of \textit{tar.gz} files; H.264 video encoding of raw video; solving \textit{mqueens} problem; Nanoscale Molecular Dynamics~(NAMD) simulation; \textit{Netflix} movie playback; execution of Random Forest (RF) machine learning algorithm; \textit{Skype} video calls; \textit{stress-ng}~\cite{stress-ng} stress test with CPU, memory, I/O, and disk workers together; playing \textit{Team Fortress 2} game; and Visual Molecular Dynamics~(VMD) modeling and visualization. \cans{It is worth mentioning~that these user-tasks represent medium to high resource-intensive tasks.}
\par
We used two different systems to build our dataset for the experiments. The configuration of these systems are as follows: (1) \textit{S1}, a laptop with an Intel Core i7-7500U @ 2.70~GHz (1 socket x 2 cores x 2 threads = 4 logical compute resources) processor, 8~GB memory, 512~GB SSD storage, NVIDIA GeForce 940MX 2~GB dedicated graphic card, \ankit{Linux~kernel~4.14} and (2) \textit{S2}, a laptop with an Intel Core i7-8550U @ 1.80~GHz (1 socket x 2 cores x 4 threads = 8 logical compute resources) processor, 16~GB memory, 512~GB SSD storage, \ankit{Linux~kernel~4.14}.
\par
\ankit{All miner programs and the \textit{perf} tool were launched in \textit{user}-mode. Even though we did not use any system-level privileges, we believe that using \textit{root} permissions for defense against cryptojacking is reasonable.} \cans{The \textit{perf} tool allows us to create per-process profile using \textit{PID}.} It is worth emphasizing that even though the dataset has been accumulated in a controlled setup, our experiments (discussed in Section~\ref{section:evaluation}) well simulate real-world scenario, where samples are collected in the real-time.

\subsection{Cryptocurrencies and miners}
\label{section:cryptocurrencies_miners}
%The miners pool their resources to combine their hashing power. In such a collaboration, miners consistently earn a portion of the block reward by generating blocks quickly. 
The probability of solving the cryptographic puzzle during mining is directly proportional to the miner's computational power/resources. Consequently, the miners pool their resources to combine their hashing power with an aim to consistently earn a portion of the block reward by solving blocks quickly. Typically, the mining pools are characterized by their hashing power. \tablename{~\ref{table:top_10_mining_pools}} shows the top-10 mining pools~\cite{top10} and the cryptocurrencies mined by them. These ten mining pools collectively constitutes the biggest share (84\% during Q1 2019) of the cryptomining business. %Please refer to \tablename{~\ref{table:acronym}} (in Appendix~\ref{appendix_a}) for the acronyms and their corresponding cryptocurrency.
\begin{table}[!htbp]
	\vspace{-2em}
	\centering
	\caption{Cryptocurrencies mined by the top-10 mining pools}
	\label{table:top_10_mining_pools}
	\resizebox{.98\columnwidth}{!}
	{
		\begin{tabular}{|c|c|c|c|c|c|c|c|c|c|c|c|c|c|c|c|c|c|}
			\hline
			\multirow{2}{*}{\textbf{N.}} & \multirow{2}{*}{\textbf{Mining pool}} & \multicolumn{16}{c|}{\textbf{Cryptocurrency}}                                                                                                                                                                                                                                                                                                                                                \\ \cline{3-18} 
			&                                       & \textbf{BCD}          & \textbf{BCH}          & \textbf{BTC}          & \textbf{BTM}          & \textbf{DASH}         & \textbf{DCR}          & \textbf{ETC}          & \textbf{ETH}          & \textbf{LTC}          & \textbf{SBTC}         & \textbf{SC}           & \textbf{UBTC}         & \textbf{XMC}          & \textbf{XMR}          & \textbf{XZC}          & \textbf{ZEC}          \\ \hline
			1                            & BTC.com                               & \xmark & \cmark & \cmark & \xmark & \xmark & \xmark & \xmark & \xmark & \xmark & \cmark & \xmark & \cmark & \xmark & \xmark & \xmark & \xmark \\ \hline
			2                            & AntPool                               & \xmark & \cmark & \cmark & \cmark & \cmark & \xmark & \cmark & \cmark & \cmark & \xmark & \cmark & \xmark & \cmark & \xmark & \xmark & \cmark \\ \hline
			3                            & ViaBTC                                & \xmark & \cmark & \cmark & \xmark & \cmark & \xmark & \cmark & \cmark & \cmark & \xmark & \xmark & \xmark & \xmark & \xmark & \xmark & \cmark \\ \hline
			4                            & SlushPool                             & \xmark & \xmark & \cmark & \xmark & \xmark & \xmark & \xmark & \xmark & \xmark & \xmark & \xmark & \xmark & \xmark & \xmark & \xmark & \cmark \\ \hline
			5                            & F2Pool                                & \xmark & \xmark & \cmark & \xmark & \cmark & \cmark & \cmark & \cmark & \cmark & \xmark & \cmark & \xmark & \cmark & \cmark & \cmark & \cmark \\ \hline
			6                            & BTC.top                               & \xmark & \cmark & \cmark & \xmark & \cmark & \xmark & \xmark & \xmark & \cmark & \xmark & \xmark & \xmark & \xmark & \xmark & \xmark & \xmark \\ \hline
			7                            & Bitclub.network                       & \xmark & \xmark & \cmark & \xmark & \cmark & \xmark & \cmark & \cmark & \xmark & \xmark & \xmark & \xmark & \xmark & \cmark & \xmark & \cmark \\ \hline
			8                            & BTCC                                  & \cmark & \cmark & \cmark & \cmark & \xmark & \xmark & \xmark & \xmark & \cmark & \cmark & \xmark & \xmark & \xmark & \xmark & \xmark & \xmark \\ \hline
			9                            & BitFury                               & \xmark & \xmark & \cmark & \xmark & \xmark & \xmark & \xmark & \xmark & \xmark & \xmark & \xmark & \xmark & \xmark & \xmark & \xmark & \xmark \\ \hline
			10                           & BW.com                                & \xmark & \xmark & \cmark & \xmark & \xmark & \xmark & \cmark & \cmark & \cmark & \xmark & \cmark & \cmark & \xmark & \xmark & \xmark & \xmark \\ \hline
		\end{tabular}
	}
\vspace{-1em}
\end{table}

\par
We considered all the cryptocurrencies mentioned in the \tablename{~\ref{table:top_10_mining_pools}} in our experiments. We used open-source miner programs to mine these cryptocurrencies. Each miner program was configured to mine with public mining pools and to utilize all available the CPUs present on the system. At the time of our experiments, the miner program for SC was not able to mine using only the CPU. Hence, we excluded SC from our experiments. To compensate SC, we included QRK whose mining algorithm - in contrast to other cryptocurrencies - uses multiple hashing algorithms. \tablename{~\ref{table:mining_algo_miners}} shows the mining algorithm of different cryptocurrencies and the CPU miners that we used.

\begin{table}[!htbp]
	\centering
	\caption{Mining algorithm and CPU miner for different cryptocurrencies}
	\label{table:mining_algo_miners}
	\resizebox{.95\columnwidth}{!}
	{
		\begin{tabular}{|c|c|c|}
			\hline
			\textbf{~Cryptocurrency~}                                         & \textbf{Mining algorithm}                                                                                                 & \textbf{CPU miner}         \\ \hline
			\canss{Bitcoin Diamond (BCD)}                                                             & X13                                                                                                                  & cpuminer-opt 3.8.8.1       \\ \hline
			\begin{tabular}[c]{@{}c@{}}\canss{Bitcoin Cash (BCH), Bitcoin (BTC)}, \\ \canss{SuperBitcoin (SBTC), UnitedBitcoin (UBTC)}\end{tabular} & SHA-256                                                                                                              & cpuminer-multi 1.3.4       \\ \hline
			\canss{Bytom (BTM)}                                                             & Tensority                                                                                                            & bytom-wallet-desktop 1.0.2 \\ \hline
			\canss{Dash (DASH)}                                                            & X11                                                                                                                  & cpuminer-multi 1.3.4       \\ \hline
			\canss{Decred (DCR)}                                                             & Blake256-r14                                                                                                         & cpuminer-multi 1.3.4       \\ \hline
			\begin{tabular}[c]{@{}c@{}}\canss{Ethereum Classic (ETC),}\\ \canss{Ethereum (ETH)}\end{tabular}                                                        & \begin{tabular}[c]{@{}c@{}}Ethash (Modified \\ Dagger-Hashimoto)\end{tabular}                                        & geth 1.7.3                 \\ \hline
				\canss{Litecoin (LTC)}                                                             & scrypt                                                                                                               & cpuminer-multi 1.3.4       \\ \hline
				\canss{Quark (QRK)}                                                            & \begin{tabular}[c]{@{}c@{}}BLAKE + Gr$\phi$stl + Blue \\ Midnight Wish + JH + \\ Keccak (SHA-3) + Skein\end{tabular} & cpuminer-multi 1.3.4       \\ \hline
				\canss{Siacoin (SC)}                                                              & BLAKE2b                                                                                                              & gominer 0.6                \\ \hline
			\canss{Monero-Classic (XMC), Monero (XMR)}                                                        & CryptoNight                                                                                                          & cpuminer-multi 1.3.4       \\ \hline
			\canss{Zcoin (XZC)}                                                             & Lyra2z                                                                                                               & cpuminer-opt 3.8.8.1       \\ \hline
			\canss{Zcash (ZEC)}                                                             & Equihash                                                                                                             & Nicehash nheqminer 0.3a    \\ \hline
		\end{tabular}
	}
\end{table}
\cans{Since our approach focuses on the underlying core PoW~algorithm, we considered one currency for every mining algorithm mentioned in \tablename{~\ref{table:mining_algo_miners}} and excluded BCH, SBTC, UBTC, ETC, and XMC in our study.} As the proof-of-concept implementation, we considered only CPU-based miner programs because each computer has at least one CPU, which cryptojackers can harness to mine. %Nevertheless, our approach is also valid to distinguish GPU-based miners because dedicated profiling tools, such as the \textit{nvprof}~\cite{nvprof} tool for NVIDIA~GPUs, allow us to monitor GPU~\textit{events}. Apart from most of the standard \textit{events} found on CPUs, GPUs have several dedicated \textit{events} that can assist in creating unique signatures for GPUs.

%\footnote{Using ccminer v2.3 on 2~GB NVIDIA GeForce MX150.}
%nvprof - CUDA Toolkit Documentation. https://tinyurl.com/z7bx3b3

\subsection{Classifier design}
\label{section:classifier_design}
In this section, we elucidate the design of our classification methodology. Algorithm~\ref{algorithm:procedure} describes the pipeline of our~classifier. \ankit{Our supervised classification algorithm begins with splitting the base-dataset of 1100 samples (2~classes x 11~instances x 50~samples) into 90-10\% stratified train-test sets, denoted as \textit{raw\_train\_set} and \textit{raw\_test\_set}.} Then, these subsets are processed as follows: 

\begin{algorithm}[!htbp]
	\caption{Pseudo code for our supervised classification.}
	\label{algorithm:procedure}
	%\hspace*{\algorithmicindent} \textbf{Input:} $S_{initial}$ 
	\begin{algorithmic}[1]
		%\renewcommand{\algorithmicrequire}{\textbf{Input:}}
		%\renewcommand{\algorithmicensure}{\textbf{Output:}}
		%\REQUIRE \ankit{in}
		%\ENSURE  \ankit{out}
		
		\FOR{each run $i$ from 1 to 10}
		\STATE Create \textit{raw\_train\_set} and \textit{raw\_test\_set} by 90-10\% stratified partitioning.
		
		\STATE \textit{Data preprocessing}\\
		$\bullet$ Replace \textit{NaN} values from \textit{raw\_train\_set} and \textit{raw\_test\_set} with arithmetic mean of the considered event.
		
		\STATE \textit{Feature engineering}\\
		$\bullet$ \textit{train\_set} := Extract\_feature(\textit{raw\_train\_set})\\
		$\bullet$ \textit{test\_set} := Extract\_feature(\textit{raw\_test\_set})
		
		\STATE \textit{Feature scaling}\\
		%We chose StandardScaler as a scaler.
		$\bullet$ scaler := StandardScaler()\\
		$\bullet$ scaler.fit(\textit{train\_set}) \hfill $\rhd$\textit{Fit scaler on \textit{train\_set}}
		
		%\STATE \textit{Transform the \textit{train\_set} and \textit{test\_set} using \textit{scaler}}\\
		$\bullet$ scaler.transform(\textit{train\_set})\\
		$\bullet$ scaler.transform(\textit{test\_set})
		
		\STATE \textit{Feature selection}\\
		$\bullet$ Compute features' importance with \textit{forests of trees} on \textit{train\_set} \cans{and select the most relevant features.} %with \textit{n\_estimators := 100} and \textit{max\_features := 50\%}
		
		\STATE \textit{Training}\\
		$\bullet$ Learn the model parameters for the given classifier (RF/SVM) on the training set using grid search with 5-fold stratified CV.
		%$\bullet$ Create and tune the model using grid search with 5-fold stratified CV.
		
		\STATE \textit{Predict/classify} the \textit{test\_set}.
		\ENDFOR			
	\end{algorithmic}
\end{algorithm}
\begin{enumerate}
	\item \textit{Data preprocessing:} 
	The first step of any machine learning-based classification is to process the raw datasets to fix any missing value. Since each event we monitor returns a numerical value, we replace the missing values, if any, with the arithmetic mean of the respective event.
	
	\item \textit{Feature engineering:} In this step, we obtain features that can be used to train a machine learning model for our prediction problem. Here, we compute 12 statistical functions (listed in \tablename{~\ref{table:features}}) for every event. This step converts each sample consisting of 300~readings (rows) x 28~\textit{events} (columns) to a single row of 336 (28~\textit{events} x 12~features) data-points. The features extracted in this phase, hereinafter referred to as \textit{train\_set} and \textit{test\_set}, are used for the subsequent stages.
	
	\begin{table}[h]
		\vspace{-2em}
		\centering
		\caption{The statistical functions that we used for our feature engineering phase}
		\label{table:features}
		%\resizebox{.98\columnwidth}{!}
		{
						\begin{tabular}{|c|c|c|}
							\hline
							0.2, 0.4, 0.6, and 0.8 quantile & 1, 2, and 3 sigma & Skewness \\ \hline
							Arithmetic and geometric mean   & Kurtosis          & Variance \\ \hline
						\end{tabular}
%			\begin{tabular}{|c|}
%				\hline
%				\textbf{Statistical function}
%				\\ \hline
%				\begin{tabular}[c]{@{}c@{}}0.2, 0.4, 0.6, and \\ 0.8 quantile\end{tabular} \\ \hline
%				1, 2, and 3 sigma                                                          \\ \hline
%				\begin{tabular}[c]{@{}c@{}}Arithmetic and \\ geometric mean\end{tabular}   \\ \hline
%				Kurtosis                                                                   \\ \hline
%				Skewness                                                                   \\ \hline
%				Variance                                                                   \\ \hline
%			\end{tabular}
		}
	\vspace{-1em}
	\end{table}
	
	\item \textit{Feature scaling:} 
	It is an essential step to eliminate the influence of large-valued features because features with larger magnitude can dominate the objective function, and thus, deterring an estimator to learn from other features correctly. Hence, we standardize features using standard scaler, which removes the mean and scale the features to unit variance.
	
	\item \textit{Feature selection:} 
	In machine learning, feature selection or dimensionality reduction is the process of selecting a subset of relevant features that are used in model construction. It aims to improve estimators' accuracy as well as to boost their performance on high-dimensional datasets.
	%To improve estimators' accuracy as well as to boost their performance on our high-dimensional datasets, we perform feature selection. 
	To do so, we calculate the importance of features using \textit{forests of trees}\ankit{~\cite{forestoftrees}} %\ankit{\footnote{\ankit{More details about calculating the importances of features here: https://tinyurl.com/y3nlad2h}}}
	\cans{and select the most relevant features.}
	
	\item \textit{Training:} %Machine learning models and parameter tuning:
	The training phase consists of learning the model parameters for the given classifier on the training set, \ie \textit{train\_set}. Given the nature of the problem, we resort to supervised machine learning procedures. In particular, we employed two of the most successful machine learning methods for classification, namely Random Forest~(RF)~\cite{Ho:1995} and Support Vector Machine~(SVM)~\cite{Cortes:1995}. 
	\par
	For model selection, we use grid search with 5-fold Cross Validation (CV). The validated hyper-parameters for RF and SVM are shown in \tablename{~\ref{table:parameters_rf}} and \tablename{~\ref{table:parameters_svm}}, respectively \cans{in Appendix~\ref{appendix_hyperparameters}}. We chose standard range of values for the hyper-parameters~\cite{HsuLibsvmTutorial2003}.
	
	\item \textit{Prediction:} Finally, prediction is made on \textit{test\_set}.
\end{enumerate}
The process is repeated ten times for a given experiment and the final results are computed over these ten runs.

\section{Evaluation}
\label{section:evaluation}
%	\section{Results}
%		\label{section:results}
We throughly evaluated our approach by performing an exhaustive set of experiments. We performed the following six different experiments: (1)~\textit{binary} classification~\cans{(Section~\ref{section:binary_classification})}; (2)~\textit{currency} classification~\cans{(Section~\ref{section:currency_classification})}; (3)~\textit{nested} classification~\cans{(Section~\ref{section:nested_classification})}; (4)~\textit{sample length}~\cans{(Section~\ref{section:sample_length})}; (5)~\textit{feature relevance}~\cans{(Section~\ref{section:feature_relevance})}; and (6)~\textit{unseen miner programs}~\cans{(Section~\ref{section:unseen_miner_programs})}. \tablename{~\ref{dataset}} describes the sample distribution in our base-dataset for each system, \ie \textit{S1} and \textit{S2}. Here, sub-classes of the mining task refer to the cryptocurrencies (discussed in Section~\ref{section:cryptocurrencies_miners}) while sub-classes of the non-mining task refer to the actual user-tasks that belong to the negative class (mentioned in Section~\ref{section:data_collection}). \ankit{We use the entire base-dataset (1100 samples per system) for each experiment, unless otherwise stated in an experiment.}
\begin{table}[!htbp]
	\centering
	\caption{Dataset: name of the task, sub-classes per task, samples per sub-class, and total samples per task for each system}
	\resizebox{.7\columnwidth}{!}
	{
		\begin{tabular}{|c|c|c|c|}
			\hline
			\textbf{Task}          & \textbf{\begin{tabular}[c]{@{}c@{}}Sub-classes\\ per task\end{tabular}} & \textbf{\begin{tabular}[c]{@{}c@{}}Samples per\\ sub-class\end{tabular}} & \textbf{\begin{tabular}[c]{@{}c@{}}Total samples\\ per task\end{tabular}} \\ \hline
			Mining                 & 11                                                                      & 50                                                                       & 550                                                                       \\ \hline
			~Non-mining~ & 11                                                                       & 50                                                                       & 550                                                                       \\ \hline
		\end{tabular}
	}
	\label{dataset}
\end{table}

\par We evaluated our classifier using standard classification metrics: Accuracy, Precision, Recall, and F$_1$~score. \cans{To increase the confidence in our results,} we report the mean and the margin of error for the results with 95\% confidence interval from ten runs of each experiment for each of the evaluation metric. %See Appendix~\ref{appendix_b} for details of these evaluation metrics and the related statistical terms. To increase the statistical significance of our results, 
\cans{We use $(\cdot)$ to indicate the best result for the metric and report the results as $mean \pm margin~of~error$.}

\subsection{Binary classification}
\label{section:binary_classification}
\cans{Our main goal is to identify whether a given instance represents the mining task or not. Hence, in this experiment, the label of each sample was defined as the positive or negative class, accordingly.} \tablename{~\ref{table:results_binary}} presents the results of the \textit{binary} classification using both RF and SVM.

\begin{table}[!htbp]
	\centering
	\caption{Results for binary classification}
	\label{table:results_binary}
	\resizebox{.8\columnwidth}{!}
	{
		\begin{tabular}{|c|c|c|c|c|c|}
			\hline
			\textbf{System}              & \textbf{Method} & \textbf{Accuracy}                & \textbf{Precision}               & \textbf{Recall}                  & \textbf{F1}                      \\ \hline
			\multirow{2}{*}{\textit{S1}} & RF     & $1.000 \pm 0.000 \bigcdot$ & $1.000 \pm 0.000 \bigcdot$ & $1.000 \pm 0.000 \bigcdot$ & $1.000 \pm 0.000 \bigcdot$ \\ \cline{2-6} 
			& SVM    & $0.999 \pm 0.002$       & $0.999 \pm 0.002$       & $0.999 \pm 0.002$       & $0.999 \pm 0.002$       \\ \hline
			\multirow{2}{*}{\textit{S2}} & RF     & $0.999 \pm 0.002 \bigcdot$ & $0.999 \pm 0.002 \bigcdot$ & $0.999 \pm 0.002 \bigcdot$ & $0.999 \pm 0.002 \bigcdot$ \\ \cline{2-6} 
			& SVM    & $0.990 \pm 0.018$       & $0.991 \pm 0.016$       & $0.990 \pm 0.018$       & $0.990\pm 0.018$        \\ \hline
		\end{tabular}
}
\end{table}
Both the RF and SVM yielded superior performance. However, RF performed better than SVM on both the systems\cans{; %the possible reason for the difference in classifiers’ performance is due to their inherent designs, by which they delineate their decision boundaries and handle outliers. 
	\canss{the possible reason for the difference in classifiers’ performance is their underlying designs - RF and SVM characterize their decision boundaries differently and also handle the outliers present in the dataset differently.} On another side, the minute variations in the performance of a given classifiers across \textit{S1} and \textit{S2} are natural and expected; mainly due to distinct dataset and data stratification.} For the sake of brevity, we report the results only for RF for the subsequent experiments. We also present the details of parameters selected by grid search in Appendix~\ref{appendix_grid_search}.

\subsection{Currency classification}
\label{section:currency_classification}
\cans{The aim of this experiment is to understand the difficulty level of classification among various cryptocurrencies. Therefore, the input dataset for this experiment contained instances only of the cryptocurrencies.} \tablename{~\ref{table:results_currency_classifier}} lists the results of the \textit{currency} classification.

\begin{figure}[!htbp]
	\centering
	\subfigure[\textit{S1}]{
		\centering
		\label{figure:confusion_matrix_gian}
		\includegraphics[trim = 2mm 2mm 2mm 6mm, clip,width=.45\linewidth]{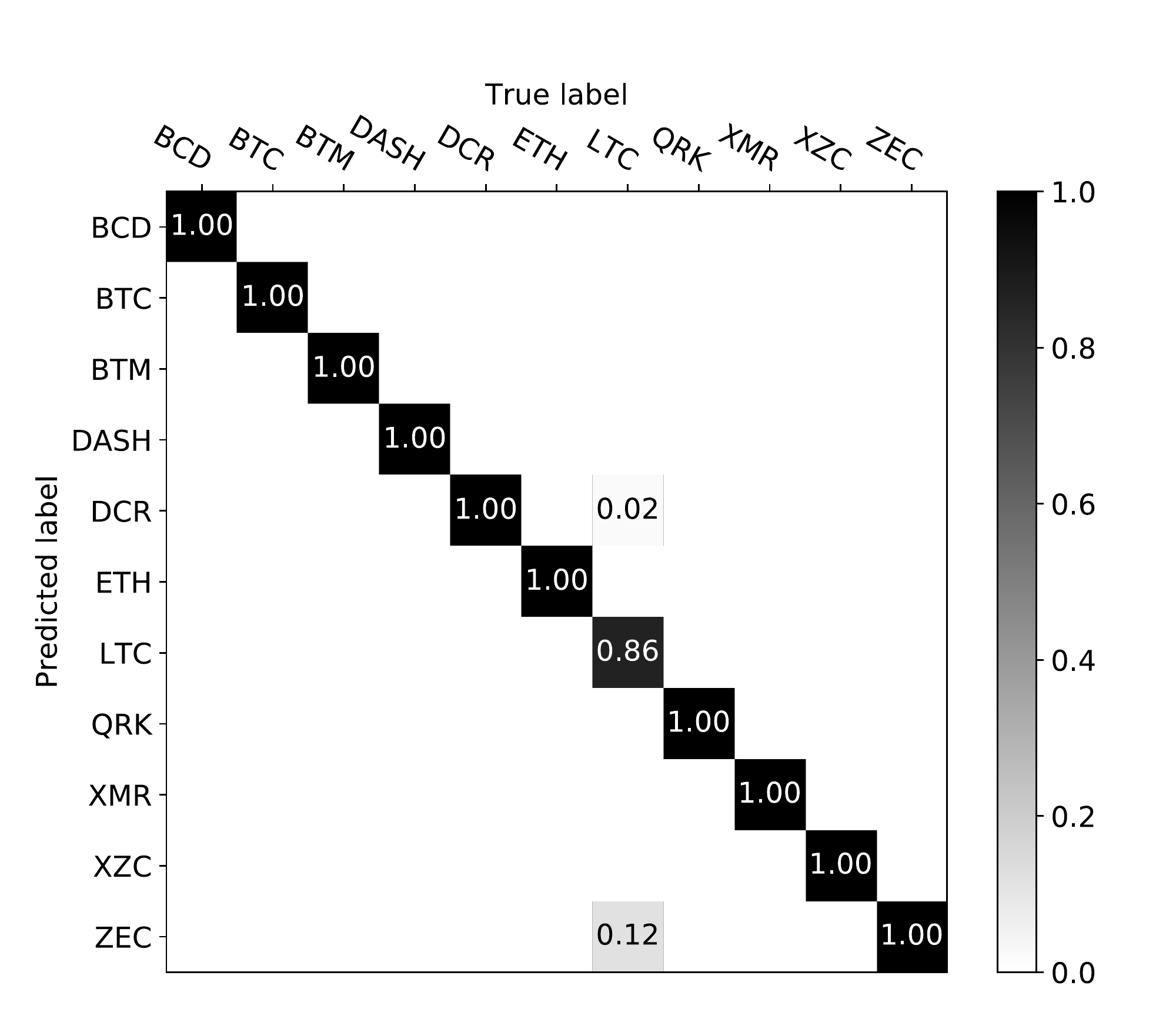}
	} 	
	\subfigure[\textit{S2}]{
		\centering
		\label{figure:confusion_matrix_samy}
		\includegraphics[trim = 2mm 2mm 2mm 6mm, clip,width=.45\linewidth]{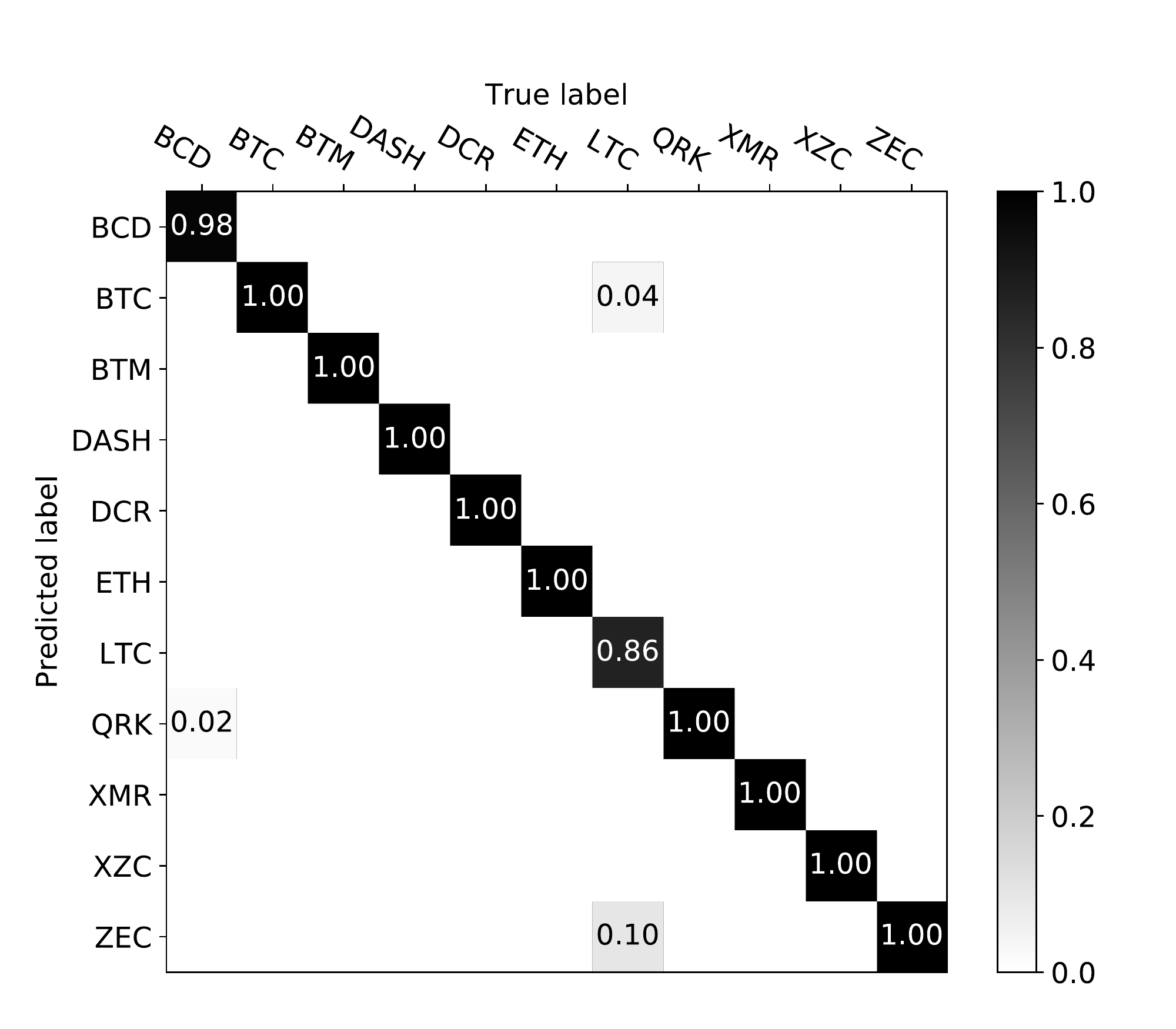}
	}
	\caption{Confusion matrix for classification among various cryptocurrencies}
	\label{figure:confusion_matrix}
\end{figure}

\begin{table}[!htbp]
	\centering
	\caption{Results for currency classification}
	\label{table:results_currency_classifier}
	\resizebox{.8\columnwidth}{!}
	{
		\begin{tabular}{|c|c|c|c|c|}
			\hline
			\textbf{System} & \textbf{Accuracy}          & \textbf{Precision}         & \textbf{Recall}            & \textbf{F1}                \\ \hline
			\textit{S1}     & $0.987 \pm 0.017$ & $0.992 \pm 0.011$ & $0.988 \pm 0.016$ & $0.985 \pm 0.020$ \\ \hline
			\textit{S2}     & $0.986 \pm 0.018$ & $0.981 \pm 0.027$ & $0.985 \pm 0.018$ & $0.982 \pm 0.024$ \\ \hline
		\end{tabular}
	}
\end{table}
\figurename{~\ref{figure:confusion_matrix}} depicts the confusion matrices for the classification among various cryptocurrencies to provide a better perception of the results. Here, \figurename{~\ref{figure:confusion_matrix_gian}} and \figurename{~\ref{figure:confusion_matrix_samy}} correspond to \textit{S1} and \textit{S2}, respectively. \ankit{The confusion matrices are drawn using the aggregate results from all the ten~runs.} \textit{Currency} classification is a multi-class classification problem, and some cryptocurrencies were misclassified among each other \ankit{(see \figurename{~\ref{figure:confusion_matrix}})}. Hence, the results are slightly lower than that of the \textit{binary}~classification. 

\subsection{Nested classification}
\label{section:nested_classification}
\cans{This experiment represents a simulation of a real-world scenario. Here, we first classify whether a given instance belongs to the positive class. If so, we identify the cryptocurrency it belongs to. Essentially, \textit{nested} classification is equivalent to performing \textit{currency} classification on the instances classified as positive in the \textit{binary} classification.} %\figurename{~\ref{fig:class-hier}} depicts the hierarchy of \textit{nested} classification.
\iffalse
\begin{figure}[!htbp]
	\centering
	\begin{tikzpicture}
	[%node distance = .8cm, auto,font=\footnotesize,
	every node/.style={node distance=.8cm},
	%comment/.style={rectangle, inner sep=3pt, text width=1.9cm, font=\scriptsize\sffamily},
	force/.style={rectangle, draw, fill=black!15, inner sep=5pt, text width=2cm, text centered, minimum height=1.1cm, font=\bfseries\footnotesize}, line width=.2mm] 
	
	% Draw forces
	\node [force] (binary) {Binary\\ classification};
	\node [force, right=2.5cm of binary] (currency) {Currency\\ classification};
	
	% Draw the links between forces
	\path[->,thick] 
	(binary) edge node[midway, above] {if `+' instance} (currency);
	%\draw [line width=0.3mm, black, -> ] (binary) edge (currency);
	\end{tikzpicture} 
	\caption{Hierarchy of nested classification}
	\label{fig:class-hier}
\end{figure}
\fi

\tablename{~\ref{table:results_nested_classifier}} shows the results of the \textit{nested} classification. In the worst case, we expect the outcome of this experiment to be lower than that of the \textit{binary} classification and \textit{currency} classification together because a crucial aspect of such staged classification is that an error made in the prediction during the primary stage influences the subsequent stage; the results for \textit{S1} shows this phenomenon. However, in a common scenario, the expected outcome of this experiment would be between the results for the \textit{binary} classification and \textit{currency} classification; the results for \textit{S2} shows this effect.

%An important aspect of such staged classification is that an error in the primary stage influences the subsequent stage.

\begin{table}[!htbp]
	\vspace{-2em}
	\centering
	\caption{Results for nested classification}
	\label{table:results_nested_classifier}
	\resizebox{.8\columnwidth}{!}
	{
		\begin{tabular}{|c|c|c|c|c|}
			\hline
			\textbf{System} & \textbf{Accuracy}          & \textbf{Precision}         & \textbf{Recall}            & \textbf{F1}                \\ \hline
			\textit{S1}     & $0.973 \pm 0.020$ & $0.972 \pm 0.026$ & $0.972 \pm 0.020$ & $0.967 \pm 0.026$ \\ \hline
			\textit{S2}     & $0.996 \pm 0.007$ & $0.997 \pm 0.006$ & $0.996 \pm 0.008$ & $0.996 \pm 0.008$ \\ \hline
		\end{tabular}
	}
\vspace{-1em}
\end{table}

%		\item \textit{Full classification:} 
%		\begin{table}[H]
%			\centering
%			\caption{Results for full classification}
%			\label{table:results_full_classifier}
%			\resizebox{.98\columnwidth}{!}
%			{
%				\begin{tabular}{|c|c|c|c|c|}
%					\hline
%					\textbf{System} & \textbf{Accuracy}          & \textbf{Precision}         & \textbf{Recall}            & \textbf{F1}                \\ \hline
%					\textit{S1}     & $1.000 \pm 0.002$ & $1.000 \pm 0.002$ & $1.000 \pm 0.002$ & $1.000 \pm 0.002$ \\ \hline
%					\textit{S2}     & $0.996 \pm 0.004$ & $0.995 \pm 0.004$ & $0.995 \pm 0.004$ & $0.995 \pm 0.004$ \\ \hline
%				\end{tabular}
%			}
%		\end{table}

\subsection{Sample length}
\label{section:sample_length}
\begin{figure}[!htbp]
	\centering
	\includegraphics[trim = 2mm 1mm 2mm 2mm, clip, width=.6\linewidth]{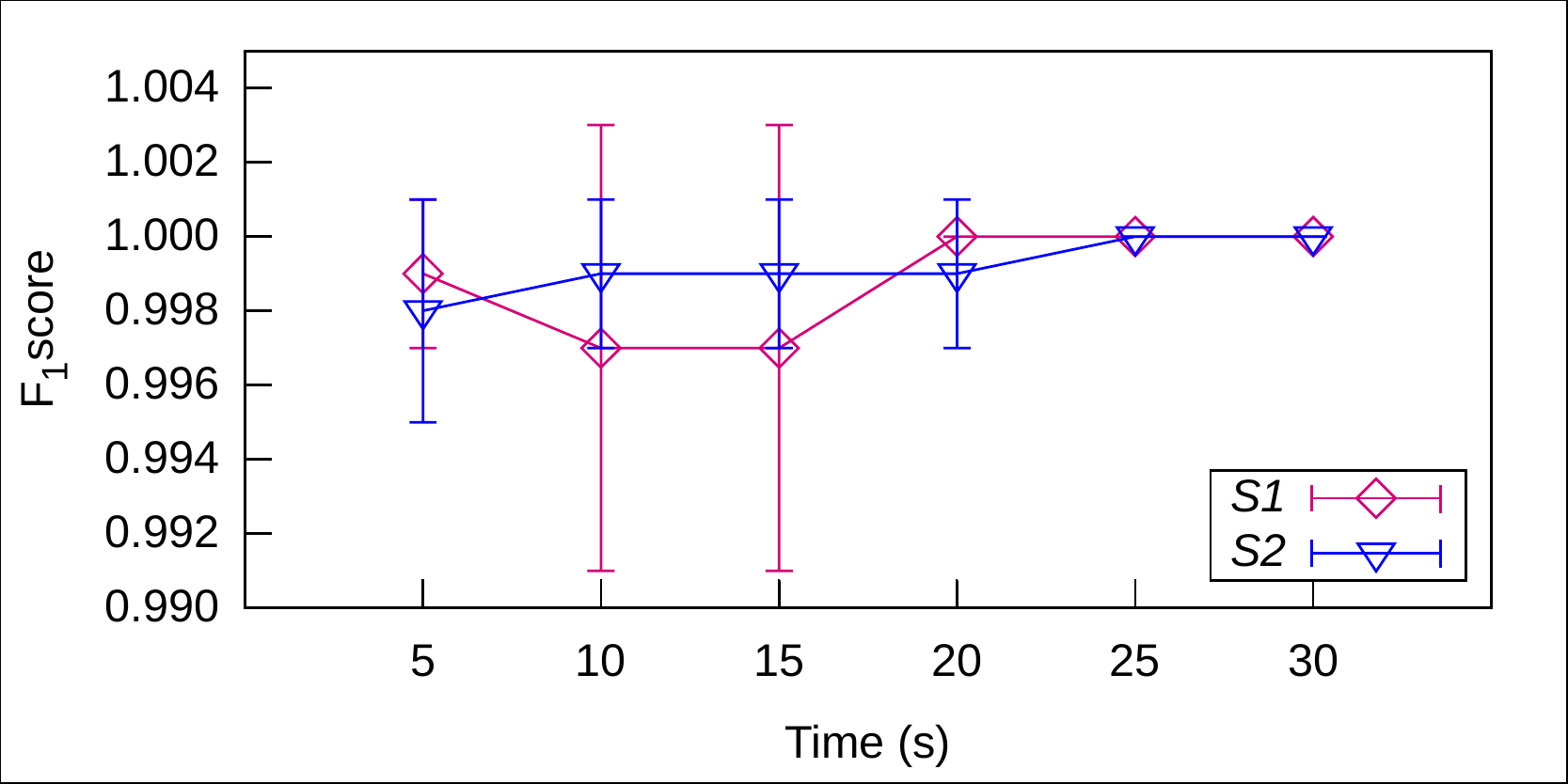}
	\caption[]{F$_1$~score for different sample lengths~(whiskers represent margin of error)}
	\label{figure:training_time}
\end{figure}

\cans{The objective of this experiment is to understand the effect of length of the samples.} For deployment in the real-world scenario, any solution - apart from being accurate - must be able to detect cryptojackers rapidly. \ankit{To this end, we performed the \textit{binary} classification of samples of a length of 5, 10, 15, 20, 25, and 30 seconds, each in separate experiments.} It is worth mentioning that we used samples of identical length for both the training and testing. \figurename{~\ref{figure:training_time}} shows the F$_1$~score when using samples of different length.

As explained in Section~\ref{section:core_concept}, the task of mining is to repeatedly execute the core PoW algorithm. Hence, even samples of shorter length can grasp the signature. As shown in \figurename{~\ref{figure:training_time}}, our system can achieve high performance with samples of 5 seconds. The dip in the curve for \textit{S1} corresponds to the thousandths digit of the F$_1$~score. For the sake of brevity, we omitted the results for sample shorter than five seconds and only focus on the required minimum sample length to attain high performance with our solution.

\subsection{Feature relevance}
\label{section:feature_relevance}
\begin{figure}[!htbp]
	\centering
	\includegraphics[trim = 2mm 1mm 2mm 2mm, clip, width=.6\linewidth]{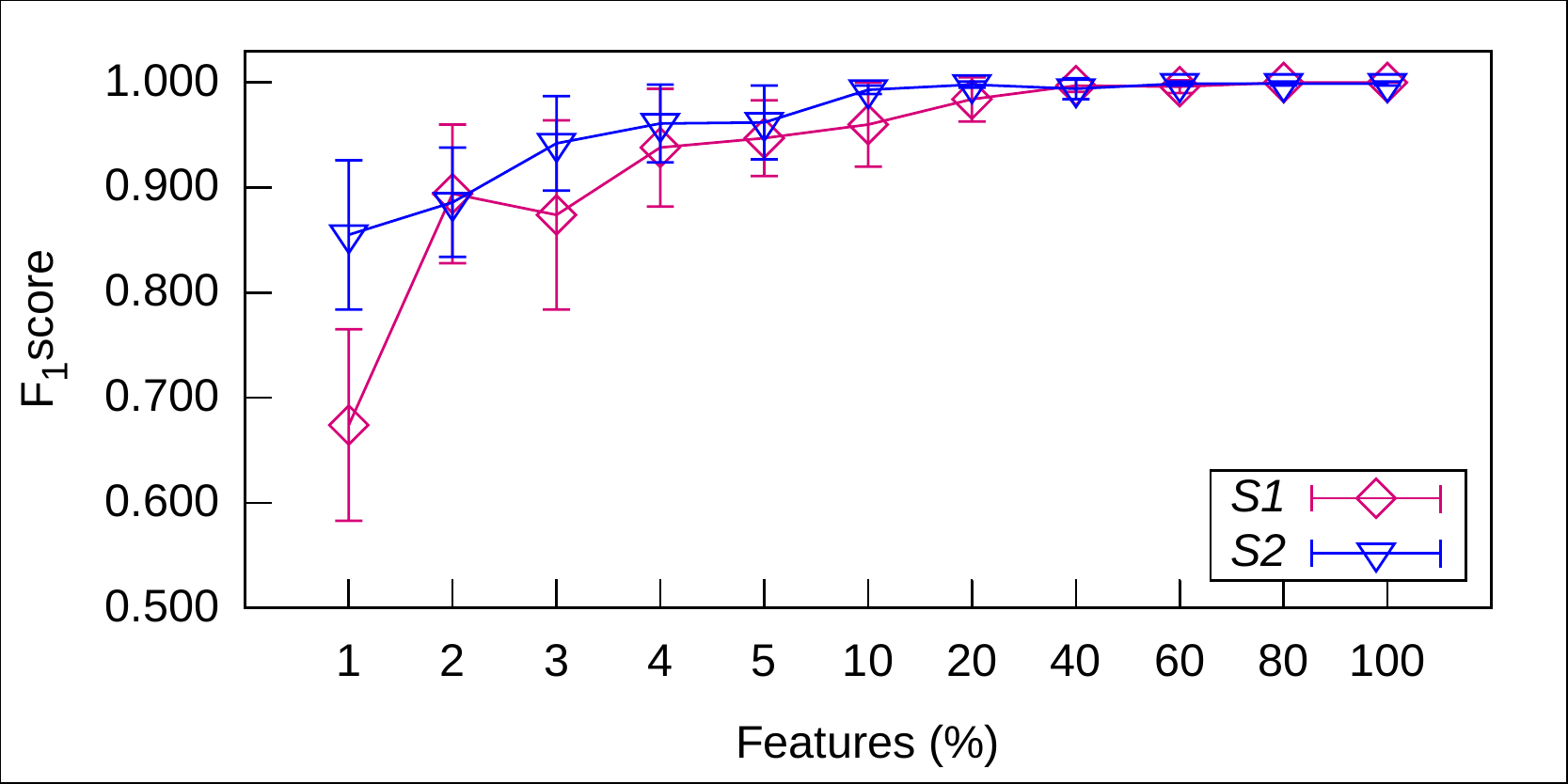}
	\caption[]{F$_1$~score for first-$\Psi$\% features~(whiskers represent margin of error)}
	\label{figure:features}
\end{figure}

%\ankit{For a successful implementation of any machine learning classifier in the real-world scenario, it is important to optimize its performance.} 
\cans{Next, we focus on our feature selection process (mentioned in Section~\ref{section:classifier_design}).} After calculating the importance of features, we sorted them in ascending order of their importance and selected the first-$\Psi$\% features to do the \textit{binary}~classification. \cans{The key idea here is to identify the lower-limit of (even less important) features required to obtain the best performance.} \figurename{~\ref{figure:features}} depicts the F$_1$~score when using first-$\Psi$\% features.

Since the features are sorted in the ascending order of their importance, \cans{we begin with the feature with lowest significance. Intuitively, including important features further improves the classification process.} As shown in the \figurename{~\ref{figure:features}}, our classifier attains high performance on both the systems using only the first-40\% (less relevant) features, which verifies/approves our feature engineering and selection process.

\subsection{Unseen miner programs}
\label{section:unseen_miner_programs}
\cans{There can be several different miner-programs available to mine a given cryptocurrency. These programs come from different developers/sources. Consequently, there can be some variations in the behavior of the miner-program itself, \eg in the code section before/after the PoW function or handling (on the programming-side) a correct nonce found while mining. The reason is that they are developed by different developers, which intuitively will cause variations. Training the model for each program may not be feasible for a variety of reasons. Hence, to investigate the effectiveness of our approach in such a situation, we set up this experiment.} Here, we selected the \textit{binary}~classification as the target where the samples from all the mining and non-mining tasks were labeled as the positive or negative class, respectively. However, we chose two additional miner programs for BTC, namely, BFGMiner~5.5 and cgminer~4.10. \ankit{ 
We collected additional 50 samples each for BFGMiner~5.5 and cgminer~4.10 on both \cans{\textit{S1} and \textit{S2}} separately. In the training phase, we used samples from one of the three miner programs for BTC. On the contrary, we used samples from one of the other two miner programs for BTC during the testing phase.} \tablename{~\ref{table:results_diff_prog_same_currency}} presents the results of classifying samples from the miner programs that were unseen in the training phase.
\begin{table}[!htbp]
	\vspace{-3em}
	\centering
	\caption{Results for unseen miner programs}
	\label{table:results_diff_prog_same_currency}
	\resizebox{.8\columnwidth}{!}
	{
		\begin{tabular}{|c|c|c|c|c|c|}
			\hline
			\textbf{System}     & \textbf{Task}     & \textbf{Accuracy} & \textbf{Precision} & \textbf{Recall}   & \textbf{F1}       \\ \hline
			\multirow{6}{*}{\textit{S1}} & $\alpha_{\beta}$  & $0.997 \pm 0.006$ & $0.997 \pm 0.006$  & $0.997 \pm 0.006$ & $0.997 \pm 0.006$ \\ \cline{2-6} 
			& $\alpha_{\gamma}$ & $0.998 \pm 0.005$ & $1.000 \pm 0.000$  & $0.997 \pm 0.006$ & $0.998 \pm 0.004$ \\ \cline{2-6} 
			& $\beta_{\alpha}$  & $1.000 \pm 0.000$ & $1.000 \pm 0.000$  & $1.000 \pm 0.000$ & $1.000 \pm 0.000$ \\ \cline{2-6} 
			& $\beta_{\gamma}$  & $0.999 \pm 0.001$ & $0.999 \pm 0.002$  & $0.999 \pm 0.002$ & $0.999 \pm 0.002$ \\ \cline{2-6} 
			& $\gamma_{\alpha}$ & $0.999 \pm 0.002$ & $0.999 \pm 0.002$  & $0.999 \pm 0.002$ & $0.999 \pm 0.002$ \\ \cline{2-6} 
			& $\gamma_{\beta}$  & $1.000 \pm 0.000$ & $1.000 \pm 0.000$  & $1.000 \pm 0.000$ & $1.000 \pm 0.000$ \\ \hline
			\multirow{6}{*}{\textit{S2}} & $\alpha_{\beta}$  & $0.999 \pm 0.001$ & $0.999 \pm 0.002$  & $0.999 \pm 0.002$ & $0.999 \pm 0.002$ \\ \cline{2-6} 
			& $\alpha_{\gamma}$ & $0.998 \pm 0.002$ & $0.997 \pm 0.003$  & $0.997 \pm 0.003$ & $0.997 \pm 0.003$ \\ \cline{2-6} 
			& $\beta_{\alpha}$  & $0.999 \pm 0.002$ & $0.998 \pm 0.003$  & $0.998 \pm 0.003$ & $0.998 \pm 0.003$ \\ \cline{2-6} 
			& $\beta_{\gamma}$  & $0.999 \pm 0.001$ & $0.999 \pm 0.002$  & $0.999 \pm 0.002$ & $0.999 \pm 0.002$ \\ \cline{2-6} 
			& $\gamma_{\alpha}$ & $0.999 \pm 0.001$ & $0.999 \pm 0.002$  & $0.999 \pm 0.002$ & $0.999 \pm 0.002$ \\ \cline{2-6} 
			& $\gamma_{\beta}$  & $0.999 \pm 0.001$ & $0.999 \pm 0.002$  & $0.999 \pm 0.002$ & $0.999 \pm 0.002$ \\ \hline
		\end{tabular}
	}\vspace{-1em}
\end{table}

\par 
The notation $X_{Y}$ means that the training was done with the samples from $X$ while the testing was done with the sample from $Y$ for BTC. Here, $\alpha~=~$cpuminer-multi~1.3.4, $\beta~=~$BFGMiner~5.5, $\gamma~=~$cgminer~4.10. It is important to mention that these results are for the classification of all the mining and non-mining tasks with BTC being trained and tested upon samples from different programs. \ankit{As discussed in Section~\ref{section:core_concept}, the miners have to execute the same core PoW algorithm for a given cryptocurrency. Hence, samples from different miner programs for a cryptocurrency retain the same signatures, which is reflected in our results.}
\par
\textit{Cross-platform classification:} \cans{Next, we evaluate the transferability of the profiles built by our approach.} We perform binary classification with additional samples from \textit{S1’} (a system with the same processor as \textit{S1}) and \textit{S2’} (a system with the same processor as \textit{S2}), and found that:~(1)~the profile of an algorithm on a given processor can be used with the help of machine learning technique to classify samples from another system with the same processor and~(2) on the contrary, the profile of an algorithm on one processor is not useful to perform classification of samples from another processor. 

%		\item \textit{Classification of unknown classes:}
%		\begin{table}[H]
%			\centering
%			\caption{Results for classification of unknown classes}
%			\label{table:results_unknown_classes}
%			\resizebox{.98\columnwidth}{!}
%			{
%				\begin{tabular}{|c|c|c|c|c|c|}
%					\hline
%					\textbf{System}     & \textbf{Task} & \textbf{Accuracy} & \textbf{Precision} & \textbf{Recall}   & \textbf{F1}       \\ \hline
%					\multirow{2}{*}{\textit{S1}} & BTC mining    & $1.000 \pm 0.004$ & $1.000 \pm 0.004$  & $1.000 \pm 0.004$ & $1.000 \pm 0.004$ \\ \cline{2-6} 
%					& h264 encoding & $1.000 \pm 0.006$ & $1.000 \pm 0.006$  & $1.000 \pm 0.006$ & $1.000 \pm 0.006$ \\ \hline
%					\multirow{2}{*}{\textit{S2}} & BTC mining    & $0.995 \pm 0.006$ & $0.995 \pm 0.006$  & $0.995 \pm 0.006$ & $0.995 \pm 0.006$ \\ \cline{2-6} 
%					& h264 encoding & $0.989 \pm 0.010$ & $0.989 \pm 0.008$  & $0.988 \pm 0.010$ & $0.988 \pm 0.010$ \\ \hline
%				\end{tabular}
%			}
%		\end{table}

\section{Limitations}
\label{section:discussion}
In this section, we address the potential limitation of our proposed approach.
\subsection{Zero-day cryptocurrencies} 
A zero-day cryptocurrency would be a currency that uses a completely new or custom PoW algorithm that was never seen before. As a matter of fact, for a cryptocurrency to obtain market value: (1)~its core-network should be supported by miners/pools; and (2)~its PoW algorithm must be accepted by the crypto-community and tested mathematically for its robustness. Therefore, the PoW algorithm for a new cryptocurrency would become public by the time it gets ready for mining, which would give us sufficient time to capture this new cryptocurrency's signature and to train our model.
\par
Importantly, miners prefer to mine cryptocurrencies that are more profitable and avoid hashing the less rewarding ones. As it happens to be, more profitable cryptocurrencies are indeed popular and their PoW algorithms are certainly known to the public. In our experiments, we considered all the popular cryptocurrencies, and our results (presented in Section~\ref{section:evaluation}) demonstrate the high quality of our proposed approach along various dimensions.

\subsection{Scalability}
The key concept of our approach is to profile the behavior of a processor's \textit{events} for mining algorithms. Since there are only a finite number of CPUs/GPUs, procuring their signature is only a matter of data collection. It might appear as a ponderous job and may be seen as a limitation of our work. But, once it is accomplished for the available CPUs/GPUs, maintaining it is relatively simpler as merely a limited number of CPUs/GPUs are released over a period of time.

\subsection{Process selection}
As mentioned in Section~\ref{section:data_collection}, our system requires per program/process-based recording of HPC for different \textit{events} as the input to the classifier. In practice, several processes run in the system. Hence, monitoring each process may consume time and can be seen as a limitation of our work. However, as shown in \figurename{~\ref{figure:training_time}}, our system can achieve high performance even with samples of 5 seconds. On another side, the miner programs attempt to use all the available resources. Therefore, an initial sorting/filtering of processes based on their resource usage can help to boost the detection process in real-time. 

\subsection{Restricted mining}
\cans{A mining strategy to evade detection from our proposed methodology can be \textit{restricted mining} that aims to change the footprint of the mining process. The essence here is that the miner program/process can be modified to perform arbitrary operations during mining. But, such maneuvers would directly affect the hashing rate and consequently the profits of mining; making the task of mining less appealing. Nevertheless, like any signature-based detection technique, it may be seen as a limitation of our work.}

\section{Conclusion and future works}
\label{section:conclusion}
Cybercriminals have developed several proficient ways to exploit cryptocurrencies with an aim to commit many unconventional financial frauds. Covert cryptomining is one of the most recent means to monetize the computational power of the victims. In this paper, we present our efficient methodology to identify covert cryptomining \cans{on users' machine}. Our solution has a broader scope \cans{- compared to the solution that are tailored to a particular cryptocurrency or a specific form (e.g., browser-based) of cryptomining on computers -} as it targets the core PoW algorithms and uses the low-performance overhead HPC that are present in modern processors to create discernible signatures. We tested our generic approach against a set of rigorous experiments that include eleven distinct cryptocurrencies. We found that our classifier attains high performance even with short samples of five seconds. 
\par
We believe that our approach is valid to distinguish GPU-based miners because dedicated profiling tools, such as the \textit{nvprof}~\cite{nvprof} tool for NVIDIA~GPUs, allow us to monitor GPU~\textit{events}. Apart from most of the standard \textit{events} found on CPUs, GPUs have several dedicated \textit{events} that can assist in creating unique signatures for GPUs. Nevertheless, we keep such investigation as part of our future work. We will also perform our experiments with a larger set of systems~(CPUs) to observe the generalization of our approach. We also hope to release a desktop application for run-time identification of covert cryptomining.
%\ankit{In the future, we will investigate the impact of samples from different operating system and virtualized environments.} We also hope to release a desktop application for run-time identification of covert cryptomining

%\section*{Acknowledgment}
%Anonymous
%Ankit Gangwal is pursuing his Ph.D. with a fellowship for international students funded by Fondazione Cassa di Risparmio di Padova e Rovigo~(CARIPARO). This work was supported in part by EU LOCARD Project under Grant H2020-SU-SEC-2018-832735, and in part by Huawei Project ``Secure Remote OTA Updates for In-Vehicle Software Systems'' under Grant HIRPO 2018040400359-2018.%This work is partially supported by the EU TagItSmart! Project (agreement H2020-ICT30-2015-688061), the EU-India REACH Project (agreement ICI+/2014/342-896), the grant n. 2017-166478 (3696) from Cisco University Research Program Fund and Silicon Valley Community Foundation, and by the grant ``Scalable IoT Management and Key security aspects in 5G systems'' from~Intel.

%\balance
\bibliographystyle{splncs04}
\bibliography{bib_short}
%\balance

%\appendices
%% if you want by leaving the argument blank
%\section{}

\begin{subappendices}
	\renewcommand{\thesection}{\Alph{section}}%
	% or try \arabic{section}
	\counterwithin{table}{section}

\section{\cans{Validated hyper-parameters}}
\label{appendix_hyperparameters}
\cans{The validated hyper-parameters for RF and SVM are shown in \tablename{~\ref{table:parameters_rf}} and \tablename{~\ref{table:parameters_svm}}, respectively.}

\begin{table}[!htbp]
	\centering
	\begin{minipage}[t]{.48\columnwidth}
		\centering
		\captionsetup{justification=centering}
		\caption{Hyper-parameters validated for RF classifier}
		\label{table:parameters_rf}
		\resizebox{.9\columnwidth}{!}
		{
			\begin{tabular}{|c|c|L}
				\hline
				\textbf{Parameter}    & \textbf{Validated values} & \textbf{Effect on the model} \\ \hline
				\textit{n\_estimators} & 
				\begin{tabular}[c]{@{}c@{}}\{10, 25, 50, 75,\\100, 125, 150\}\end{tabular} &		\fixcolumn{Number of trees use in the ensemble.} \\ \hline
				\textit{max\_depth}    & [2, $ \infty$)               &			\fixcolumn{Maximum depth of the trees.} \\ \hline
				\textit{max\_features} & `auto', `log2'          &				\fixcolumn{Number of features to consider when looking for the best split.}                            \\ \hline
				\textit{split\_criterion}    & \textit{gini, entropy}           &		\fixcolumn{Criterion used to split a node in a decision tree.}                           \\ \hline
				\textit{bootstrap}    & \textit{true, false}             &		\fixcolumn{Bootstrap Aggregation (a.k.a. bagging) is a technique that reduces model variances (overfitting) and improves the outcome of learning on limited sample or unstable datasets.}                           \\ \hline
				\textit{random\_state} & 10                        &			\fixcolumn{The seed used by the random number generator.}                           \\ \hline
			\end{tabular}
		}
		
	\end{minipage}%
	\begin{minipage}[t]{.48\columnwidth}
		\centering
		\captionsetup{justification=centering}
		\caption{Hyper-parameters validated for SVM classifier}
		\label{table:parameters_svm}
		\resizebox{.9\columnwidth}{!}
		{
			\begin{tabular}{|c|c|L}
				\hline
				\textbf{Parameter}       & \textbf{Validated values}             & \textbf{Effect on the model} \\ \hline
				\textit{kernel}        & 
				\begin{tabular}[c]{@{}c@{}}`rbf', `poly',\\`sigmoid'\end{tabular} &				\fixcolumn{Specifies the kernel type to be used in the algorithm.}                              \\ \hline
				\textit{C}               & [$10^{-3}$, $10^5$]        &					\fixcolumn{Regularization parameter that controls the trade-off between the achieving a low training error and a low testing error that is the ability to generalize your classifier to unseen data.}                              \\ \hline
				\textit{$\gamma$}         & 
				
				\begin{tabular}[c]{@{}c@{}}`auto',\\~[$10^{-7}$, $10^{3}$]\end{tabular} &	\fixcolumn{Shape parameter of the RBF kernel which defines how an example influence in the final classification.}                              \\ \hline
				\textit{degree}        & default=3                       &				\fixcolumn{Degree of the polynomial kernel function (`poly'). Ignored by all other kernels.}                              \\ \hline
				\textit{random\_state} & 10                                    &		\fixcolumn{The seed of the pseudo random number generator used when shuffling the data for probability estimates.}                              \\ \hline
			\end{tabular}
		}
	\end{minipage} 
\end{table}

\section{Parameters selected by grid search}
\label{appendix_grid_search}
% \usepackage{multirow}
Here, we list the frequency of paramenter-values selected by grid search over ten-runs of different experiments. \tablename{~\ref{table:grid_search_svm}} corresponds to \textit{binary} classification experiment with SVM while \tablename{~\ref{table:grid_search_rf}} corresponds to \textit{binary}, \textit{currency}, and \textit{full}~classification experiments with RF for both \textit{S1} and \textit{S2}.
\begin{table}[!htbp]
	\centering
	\caption{\textit{Binary} classification with SVM}
	\label{table:grid_search_svm}
	\resizebox{.6\columnwidth}{!}
	{
	\begin{tabular}{|c|c|c|c|}
		\hline
		\textbf{Parameter} & \textbf{Value} & \textbf{\begin{tabular}[c]{@{}c@{}}N. of times\\ selected on  \textit{S1}\end{tabular}} & \textbf{\begin{tabular}[c]{@{}c@{}}N. of times\\ selected on  \textit{S2}\end{tabular}} \\ \hline
		\multirow{3}{*}{\textit{kernel}} & `rbf' & 7 & 6 \\ \cline{2-4} 
		& `poly' & 1 & 0 \\ \cline{2-4} 
		& `sigmoid' & 2 & 4 \\ \hline
		\multirow{6}{*}{\textit{C}} & 0.01 & 1 & 0 \\ \cline{2-4} 
		& 0.1 & 0 & 4 \\ \cline{2-4} 
		& 1 & 1 & 1 \\ \cline{2-4} 
		& 10 & 3 & 2 \\ \cline{2-4} 
		& 100 & 2 & 2 \\ \cline{2-4} 
		& 1000 & 3 & 1 \\ \hline
		\multirow{5}{*}{\textit{$\gamma$}} & 0.0001 & 2 & 1 \\ \cline{2-4} 
		& 0.001 & 1 & 4 \\ \cline{2-4} 
		& 0.01 & 2 & 1 \\ \cline{2-4} 
		& 0.1 & 2 & 0 \\ \cline{2-4} 
		& `auto' & 3 & 4 \\ \hline
	\end{tabular}
}
\end{table}

% \usepackage{multirow}
\begin{table}[H]
	\centering
	\caption{Different classifications with RF}
	\label{table:grid_search_rf}
	\resizebox{.98\columnwidth}{!}
	{
		\begin{tabular}{cc|c|c|c|c|c|c|}
			\cline{3-8}
			&  & \multicolumn{2}{c|}{\textit{Binary} classification} & \multicolumn{2}{c|}{\textit{Currency} classification} & \multicolumn{2}{c|}{\textit{Full} classification} \\ \hline
			\multicolumn{1}{|c|}{\textbf{Parameter}} & \textbf{Value} & \textbf{\begin{tabular}[c]{@{}c@{}}N. of times\\ selected on  \textit{S1}\end{tabular}} & \textbf{\begin{tabular}[c]{@{}c@{}}N. of times\\ selected on  \textit{S2}\end{tabular}} & \textbf{\begin{tabular}[c]{@{}c@{}}N. of times\\ selected on  \textit{S1}\end{tabular}} & \textbf{\begin{tabular}[c]{@{}c@{}}N. of times\\ selected on  \textit{S2}\end{tabular}} & \textbf{\begin{tabular}[c]{@{}c@{}}N. of times\\ selected on  \textit{S1}\end{tabular}} & \textbf{\begin{tabular}[c]{@{}c@{}}N. of times\\ selected on  \textit{S2}\end{tabular}} \\ \hline
			\multicolumn{1}{|c|}{\multirow{2}{*}{\textit{bootstrap}}} & \textit{true} & 10 & 10 & 10 & 10 & 10 & 10 \\ \cline{2-8} 
			\multicolumn{1}{|c|}{} & \textit{false} & 0 & 0 & 0 & 0 & 0 & 0 \\ \hline
			\multicolumn{1}{|c|}{\multirow{2}{*}{\textit{max\_features}}} & `log2' & 3 & 4 & 5 & 3 & 5 & 1 \\ \cline{2-8} 
			\multicolumn{1}{|c|}{} & `auto' & 7 & 6 & 5 & 7 & 5 & 9 \\ \hline
			\multicolumn{1}{|c|}{\multirow{6}{*}{\textit{max\_depth}}} & 2 & 0 & 0 & 4 & 1 & 0 & 0 \\ \cline{2-8} 
			\multicolumn{1}{|c|}{} & 3 & 5 & 5 & 5 & 5 & 5 & 1 \\ \cline{2-8} 
			\multicolumn{1}{|c|}{} & 4 & 2 & 1 & 0 & 3 & 4 & 7 \\ \cline{2-8} 
			\multicolumn{1}{|c|}{} & 5 & 2 & 2 & 1 & 1 & 1 & 2 \\ \cline{2-8} 
			\multicolumn{1}{|c|}{} & 6 & 1 & 0 & 0 & 0 & 0 & 0 \\ \cline{2-8} 
			\multicolumn{1}{|c|}{} & 7 & 0 & 2 & 0 & 0 & 0 & 0 \\ \hline
			\multicolumn{1}{|c|}{\multirow{2}{*}{\textit{split\_criterion}}} & \textit{gini} & 9 & 9 & 10 & 6 & 10 & 10 \\ \cline{2-8} 
			\multicolumn{1}{|c|}{} & \textit{entropy} & 1 & 1 & 0 & 4 & 0 & 0 \\ \hline
			\multicolumn{1}{|c|}{\multirow{7}{*}{\textit{n\_estimators}}} & 10 & 2 & 3 & 0 & 5 & 0 & 0 \\ \cline{2-8} 
			\multicolumn{1}{|c|}{} & 25 & 5 & 1 & 1 & 2 & 1 & 0 \\ \cline{2-8} 
			\multicolumn{1}{|c|}{} & 50 & 2 & 1 & 4 & 1 & 0 & 1 \\ \cline{2-8} 
			\multicolumn{1}{|c|}{} & 75 & 0 & 0 & 2 & 2 & 0 & 0 \\ \cline{2-8} 
			\multicolumn{1}{|c|}{} & 100 & 0 & 0 & 2 & 0 & 5 & 5 \\ \cline{2-8} 
			\multicolumn{1}{|c|}{} & 125 & 1 & 4 & 0 & 0 & 3 & 1 \\ \cline{2-8} 
			\multicolumn{1}{|c|}{} & 150 & 0 & 1 & 1 & 0 & 1 & 3 \\ \hline
		\end{tabular}
	}
\end{table}
\end{subappendices}

%\enlargethispage{\baselineskip}
\balance
\end{document}